\documentclass[a4paper,12pt]{article}

\usepackage{hyperref}
\usepackage{url}
\usepackage{amssymb}
\usepackage[footnotesize]{caption}
\usepackage{graphicx}
\usepackage[font=scriptsize]{subcaption}
\usepackage[dvipsnames]{xcolor}
\usepackage{mathtools}
\usepackage{braket}
\usepackage{cite}
\usepackage{multirow}
\usepackage{relsize}
\usepackage{fullpage}
\usepackage{makecell}
\usepackage{blkarray}
\usepackage{placeins}
\usepackage[utf8]{inputenc}
\usepackage{empheq}

\numberwithin{equation}{section}
\newcommand{\overbar}[1]{\mkern 1.5mu\overline{\mkern-1.5mu#1\mkern-1.5mu}\mkern 1.5mu}
\newcommand{\secheadmath}[1]{\texorpdfstring{$#1$}{TEXT}} 
\newcommand{\pmatr}[1]{\begin{pmatrix} #1 \end{pmatrix}}
\newcommand{\simlt}{~\mbox{\smaller\(\lesssim\)}~}
\newcommand{\simgt}{~\mbox{\smaller\(\gtrsim\)}~}

\newcommand{\exto}[1]{{\smaller$ \times 10^{#1} $}}
\newcommand{\degr}{$^\circ$}

\begin{document}
\begin{titlepage}
\setcounter{page}{0}
\vspace*{0.7cm}

\begin{center}
{\bf\Large Towards a complete $\bf \Delta(27) \times SO(10)$ SUSY GUT
} \\[12mm]
Fredrik~Bj\"{o}rkeroth$^{\star}$%
\footnote{E-mail: {\tt f.bjorkeroth@soton.ac.uk}},
Francisco~J.~de~Anda$^{\dagger}$%
\footnote{E-mail: \texttt{franciscojosedea@gmail.com}},
Ivo~de~Medeiros~Varzielas$^{\star}$%
\footnote{E-mail: \texttt{ivo.de@soton.ac.uk}},
Stephen~F.~King$^{\star}$%
\footnote{E-mail: \texttt{king@soton.ac.uk}}
\\[-2mm]

\end{center}
\vspace*{0.50cm}
\centerline{$^{\star}$ \it
School of Physics and Astronomy, University of Southampton,}
\centerline{\it 
SO17 1BJ Southampton, United Kingdom }
\vspace*{0.2cm}
\centerline{$^{\dagger}$ \it
Departamento de F{\'i}sica, CUCEI, Universidad de Guadalajara, M{\'e}xico}
\vspace*{1.20cm}

\begin{abstract}
{\noindent
We propose a renormalisable model based on $\Delta(27)$ family symmetry with an $SO(10)$ grand unified theory (GUT) leading to a novel form of spontaneous geometrical CP violation.
The symmetries, including $\Delta(27)$ and $\mathbb{Z}_{9} \times \mathbb{Z}_{12} \times \mathbb{Z}_{4}^{R}$,
are broken close to the GUT breaking scale to yield
the minimal supersymmetric standard model (MSSM) with the standard R-parity.
$SO(10)$ is broken via $SU(5)$ with doublet-triplet splitting achieved by a version of the Dimopoulos-Wilczek (missing VEV) mechanism.
Low-scale Yukawa structure is dictated by the coupling of matter to $ \Delta(27) $ antitriplets $ \overbar{\phi} $ whose VEVs are aligned in the CSD3 directions by the superpotential.
Light physical Majorana neutrinos masses emerge from a specific implementation of the seesaw mechanism within $SO(10)$. The model predicts a normal neutrino mass hierarchy with the 
best-fit lightest neutrino mass between $0.32-0.38$ meV, CP-violating oscillation phase $\delta^l\approx (275-280)^{\circ}$ and the remaining neutrino parameters all within $1 \sigma$ of their best-fit experimental values.
}
\end{abstract}
\end{titlepage}

\section{Introduction}

The Standard Model (SM) cannot possibly be a complete theory, since it does not provide an explanation for neutrino mass and mixing. In addition, it provides no glimmer of insight into the flavour and CP puzzles, or the origin of three distinct gauge forces. A very ambitious approach, capable in principle of addressing all these questions, is the idea of a Grand Unified Theory (GUT) combined with 
a family symmetry which can control the structure of the Yukawa couplings, leading to a predictive theory 
of flavour. In addition, Supersymmetry (SUSY) is the most elegant way to ensure gauge coupling
unification, also stabilising the Higgs mass (for a review see e.g. \cite{King:2013eh}).
The state of the art is to combine a realistic GUT (addressing issues like doublet-triplet splitting) with predictive flavour structures \cite{Antusch:2009gu, Antusch:2014poa}, and we proposed a fairly complete $A_4\times SU(5)$ SUSY GUT of flavour
along these lines \cite{Bjorkeroth:2015ora}.
However the most ambitious, but also the most challenging, of such theories are those based on 
$SO(10)$ \cite{Fritzsch:1974nn} where three right-handed neutrinos are predicted and neutrino mass is therefore
inevitable. Typically such theories are very difficult to reconcile with a family symmetry,
and generally involve rather large dimensional Higgs representations. 

In this paper we propose a realistic and fairly complete model, capable of addressing all the above questions unanswered by the SM, based on  $\Delta(27) \times  SO(10)$
with a CP symmetry at the high scale.
The choice of $\Delta(27)$ is primarily due to its triplet and anti-triplet representation, such that there is no invariant between two triplets, which is convenient due to the SM fermions being placed all in a single $SO(10)$ $ \mathbf{16} $, $\Delta(27)$ triplet. 
In addition, the non-trivial singlets of $\Delta(27)$ are also useful, as they are used to give rise CP violating phases that are related to the group rather than arbitrary parameters in the Lagrangian. We therefore describe this as spontaneous geometrical CP violation \cite{Branco:1983tn}, in this model in a novel form, as it fixes relative of phases between distinct flavons.
The model has many attractive features, including the use of only the lower dimensional ``named'' representations of $SO(10)$, i.e. the singlet, fundamental, spinor or adjoint representations.
$SO(10)$ is broken via $SU(5)$ with doublet-triplet splitting achieved by a version of the Dimopoulos-Wilczek (DW) or missing VEV mechanism \cite{DW}.

The renormalisable $\Delta(27) \times  SO(10)$ 
model also involves a discrete $\mathbb{Z}_{9} \times \mathbb{Z}_{12} \times \mathbb{Z}_{4}^{R}$. 
The family symmetries are broken close to the GUT breaking scale to yield
the minimal supersymmetric standard model (MSSM) supplemented by a right-handed neutrino seesaw mechanism \cite{Minkowski:1977sc, Ramond:1979py}, where the $\mathbb{Z}_{4}^{R}$ is the origin of the MSSM R-parity \cite{Lee:2011dya}. 
The model is realistic in the sense that it provides a successful (and natural) description of the quark and lepton (including neutrino) mass and mixing spectra, including spontaneous CP violation. The low-scale Yukawa structure is dictated by the coupling of matter to $ \Delta(27) $ antitriplets $ \overbar{\phi} $ whose VEVs are aligned in the CSD3 directions by a superpotential.
Light physical Majorana neutrinos masses emerge from a specific implementation of the seesaw mechanism within $SO(10)$.
It is fairly complete in the sense that GUT and family symmetry breaking are addressed, including doublet-triplet splitting and the origin of the MSSM $\mu$ term.  

We emphasise the predictive nature of the model. 
Large lepton mixing is accounted for by the seesaw mechanism \cite{Minkowski:1977sc} with constrained sequential dominance (CSD) \cite{King:1998jw}. 
The basic goal of the flavour sector in these models is to couple the SM fermions to flavons $ \overbar{\phi}_{\rm atm} $, $ \overbar{\phi}_{\rm sol} $ and $ \overbar{\phi}_{\rm dec} $, whose VEVs are aligned in the CSD3 direction \cite{King:2013iva,Bjorkeroth:2014vha},%
\footnote{CSD4 models have been discussed in \cite{King:2013hoa}.}
i.e. where
\begin{equation}
	\overbar{\phi}_{\rm atm} \sim \pmatr{0\\1\\1} , \qquad \overbar{\phi}_{\rm sol} \sim \pmatr{1\\3\\1} , \qquad \overbar{\phi}_{\rm dec} \sim \pmatr{0\\0\\1}.
	\label{eq:csd3alignments}
\end{equation}
We achieve this in a way that is compatible with an $ SO(10) $ GUT, i.e. where all fermion states may be united in a \textbf{16} of $ SO(10) $, and left- and right-handed fermions transform equally under the family symmetry. Since $ SO(10) $ constrains the Dirac couplings of all leptons and quarks to be equal (within a family), 
it is actually rather non-trivial that the successful scheme in the lepton sector will translate to success in the quark sector. Remarkably we find that we can attain good fits to data for quark and lepton masses, mixings and phases. 
This is notably different from our previous work \cite{Bjorkeroth:2015ora} based on $ SU(5) $ with CSD3, wherein the three generations of fermions were not all unified into triplets of the family symmetry.

The full literature on flavoured SUSY GUTs \cite{Ramond:1979py}, i.e. which involve
a family symmetry, is quite extensive (for an incomplete list see e.g. \cite{huge,SO10}), but there have been relatively few attempts in the literature to combine an $ SO(10) $ GUT with 
a discrete non-Abelian family symmetry \cite{SO10} and we would argue that none are as successful or complete as the present one.
The goal of all these models is clear: to address the questions left unanswered by the SM. It will take some time and (experimental) effort to resolve these models. However the most promising models are those that make testable predictions while being theoretically complete and consistent. 

The layout of the remainder of the paper is as follows.
In Section \ref{sec:yukawa} we present a renormalisable Yukawa superpotential and discuss how it leads to the fermion mass matrices.
In Section \ref{sec:alignment} we show how the CSD3 alignment is produced in $ \Delta(27) $, how the flavon VEVs are driven, and their relative phases fixed.
In Section \ref{sec:gutbreaking} we show how $ SO(10) $ is broken, and how we achieve doublet-triplet splitting.
In Section \ref{sec:fit} we give a numerical fit of model parameters to the masses and mixing parameters as given by data.
Section \ref{sec:conclusion} concludes.

\section{A model based on \secheadmath{\Delta(27)\times SO(10)} with CSD3}
\label{sec:yukawa}
\subsection{Yukawa superpotential and field content}

The most important field content is given in Table \ref{tab:funfields}. In Table \ref{tab:fermionmess} we have the messengers with $ R $-charge 1 which result in the superpotential in Eq.~\ref{eq:sYW}. Higgs fields are typically denoted by their $ SO(10) $ representation, with two $ \mathbf{10} $s that couple respectively to the up-type and down-type MSSM fields at the low scale. The fields $ \overbar{\phi}_i $ are flavons that are antitriplets under $ \Delta(27) $, and are named in accordance with their respective roles in the CSD3 scheme. The messenger fields are typically indexed by their $ \mathbb{Z}_9 $ charge, while each prime tick corresponds to an additive $ \mathbb{Z}_{12} $ charge of 3.

\begin{table}
\centering
\footnotesize
\begin{minipage}[b]{0.4\textwidth}
\captionsetup{width=0.9\textwidth}
\begin{tabular}{| c | c@{\hskip 5pt}c | c c c |}
\hline
\multirow{2}{*}{\rule{0pt}{4ex}Field}	& \multicolumn{5}{c |}{Representation} \\
\cline{2-6}
\rule{0pt}{3ex}			& $\Delta(27)$ & $SO(10)$ & $\mathbb{Z}_{9}$ &$\mathbb{Z}_{12}$ & $\mathbb{Z}_4^R$ \\ [0.75ex]
\hline \hline
\rule{0pt}{3ex}%
$\Psi$ 			& 3 & 16 & 0 & 0 &1\\
\rule{0pt}{3ex}%
$H_{10}^u$ & 1 & 10 & $6$ & 0 & 0\\
$H_{10}^d$ & 1 & 10 & $5$ & 0& 0\\
$H_{45}$ & 1 & 45 & 0 & 0&0\\
$H_{45}'$ & 1 & 45 & 0& 3 &0\\
$H_{DW}$& 1 & 45 & $6$ & 0 &2\\
$Z$& 1 & 1 & 0& 0 &2\\
$Z''$& 1 & 1 & 0& 6 &2\\
$H_{\overbar{16}}$ & 1 & $\overbar{16}$ & $6$ & 0&0\\
$H_{16}$ & 1 & $16$ & $2$ & 0&2\\
\rule{0pt}{3ex}%
$\overbar{\phi}_{\rm dec} $ &  $ \overbar{3} $ & 1 & $6$ & 0&0\\ 
$\overbar{\phi}_{\rm atm} $ &  $ \overbar{3} $ & 1 & $1$ & 0 &0\\ 
$\overbar{\phi}_{\rm sol} $ &  $ \overbar{3} $ & 1 & $5$ & 6&0\\ 
$\xi$ & 1 & 1 & $1$ & 0&0\\[0.5ex]
\hline
\end{tabular}
\caption{Superfields important for quark and lepton Yukawa couplings.}
\label{tab:funfields}
\end{minipage}%
\qquad\begin{minipage}[b]{0.53\textwidth}
\captionsetup{width=0.9\textwidth}
\centering
\begin{tabular}{| c | c@{\hskip 3pt}c | c c c |}
\hline
\multirow{2}{*}{\rule{0pt}{4ex}Field}	& \multicolumn{5}{c |}{Representation} \\
\cline{2-6}
\rule{0pt}{3ex}			& $\Delta(27)$ & $SO(10)$ & $\mathbb{Z}_{9}$ &$\mathbb{Z}_{12}$ & $\mathbb{Z}_4^R$ \\ [0.75ex]
\hline \hline
\rule{0pt}{3ex}%
$ \chi_i $ & 1 & 16 & $ i \in \{1,5,6,7\} $ & 0 & 1 \\
$ \overbar{\chi}_i $ & 1 & $ \overbar{16} $ & $ i \in \{8,4,3,2\} $ & 0 & 1 \\
\rule{0pt}{3ex}%
$ \chi^{\prime\prime}_i $ & 1 & 16 & $ i \in \{5,6,7\} $ & 6 & 1 \\
$ \overbar{\chi}^{\prime\prime}_i $ & 1 & $ \overbar{16} $ & $ i \in \{4,3,2\} $ & 6 & 1 \\
\rule{0pt}{3ex}%
$ \chi^{\prime}_6 $ & 1 & 16 & 6 & 3 & 1 \\
$ \overbar{\chi}^{\prime\prime\prime}_3 $ & 1 & $ \overbar{16} $ & 3 & 9 & 1 \\
\rule{0pt}{3ex}%
$ \Omega_i $ & 1 & 1 & $ i \in \{0,...\,,8\} $ & 0 & 1 \\
$ \Omega_i^{\prime\prime} $ & 1 & 1 & $ i \in \{3,4,5,6\} $ & 6 & 1 \\[0.5ex]
\hline
\end{tabular}
\caption{Messengers with unit R-charge.\vspace{2.8ex}}
\label{tab:fermionmess}
\end{minipage}%
\end{table}

The MSSM matter content is collected in $ \Psi $, a $ \mathbf{16} $ of $ SO(10) $ and a triplet under $ \Delta(27) $.
The two Higgs doublets arise from $ H^u_{10} $ and $ H^d_{10} $, both $ \mathbf{10} $ representations of $ SO(10) $, where one only gets a VEV in the ($ SU(2) $) $H_u$ direction and the other in the $H_d$ direction. If we didn't have the two $H_{10}$ we would get the erroneous relation
\begin{equation}
\tan \beta ~m_{ij}^d=m^u_{ij},
\label{eq:tanwrong}
\end{equation}
which gives no CKM mixing.
The $H_{\overbar{16}}$ breaks $SO(10)\to SU(5)$ and gives masses to right-handed neutrinos.

The $H_{45}$ obtains a VEV that breaks $SU(5)$ to the Standard Model group, i.e. $ SU(5) \to SU(3) \times SU(2) \times U(1) $.
It also gives the necessary Clebsch-Gordan coefficients to give the correct masses. 
Since it has no $\mathbb{Z}$ charge and the messengers should be in the $ \mathbf{16}$ representation, they can have a renormalizable mass or a mass depending on the VEV of the $ \mathbf{45}$. This is discussed further in Section \ref{sec:diagramsandcg}.

The Yukawa superpotential that produces the quark and lepton mass matrices is
\begin{equation}
\medmuskip=3mu
\thinmuskip=0mu
\thickmuskip=0mu
\begin{split}
	\mathcal{W}_Y &~=~\Psi_i \Psi_j H_{10}^u 
	\left[ \overbar{\phi}^{i}_{\rm dec}\overbar{\phi}^{j}_{\rm dec} \sum_{n=0}^2 \frac{\lambda^{(u)}_{{\rm dec},n}}{\braket{H_{45}}^n M_\chi^{2-n}}
	+ \overbar{\phi}^{i}_{\rm atm} \overbar{\phi}^{j}_{\rm atm} \xi \sum_{n=0}^3 \frac{\lambda^{(u)}_{{\rm atm},n}}{\braket{H_{45}}^{n} M_\chi^{3-{n}}} \right.
	\\
	& \left. \hspace{14ex} + \overbar{\phi}^{i}_{\rm sol} \overbar{\phi}^{j}_{\rm sol} \xi^2 \sum_{n=0}^4 \frac{\lambda^{(u)}_{{\rm sol},n}}{\braket{H_{45}}^n M_\chi^{4-n}} +\overbar{\phi}^{i}_{\rm sol}\overbar{\phi}^{j}_{\rm dec}\xi 
	\bigg(\frac{\lambda^{(u)}_{{\rm sd},1}}{\braket{H_{45}'}^2 M_\chi}+\frac{\lambda^{(u)}_{{\rm sd},2}}{\braket{H_{45}'}^2 \braket{H_{45}}}\bigg) \right] 
	\\
	&\quad+ \Psi_i \Psi_j H_{10}^d 
	\left[ \overbar{\phi}^{i}_{\rm dec} \overbar{\phi}^{j}_{\rm dec}\xi \sum_{n=0}^3 \frac{\lambda^{(d)}_{{\rm dec},n}}{\braket{H_{45}}^n M_\chi^{3-n}}
	+ \overbar{\phi}^{i}_{\rm atm} \overbar{\phi}^{j}_{\rm atm} \xi^2 \sum_{n=0}^4 \frac{\lambda^{(d)}_{{\rm atm},n}}{\braket{H_{45}}^{n} M_\chi^{4-{n}}} \right.\\
	&  \left. \hspace{14ex} + \overbar{\phi}^{i}_{\rm sol} \overbar{\phi}^{j}_{\rm sol} \xi^3 \sum_{n=0}^5 \frac{\lambda^{(d)}_{{\rm sol},n}}{\braket{H_{45}}^n M_\chi^{5-n}} \right] \\
	&\quad+ \Psi_i \Psi_j H_{\overbar{16}}H_{\overbar{16}} 
	\left[
	\overbar{\phi}^{i}_{\rm dec} \overbar{\phi}^{j}_{\rm dec} \xi^3\frac{\lambda^{(M)}_{{\rm dec}}}{M_\chi^2 M_{\Omega_{\rm dec}}^4} 
	+ \overbar{\phi}^{i}_{\rm atm} \overbar{\phi}^{j}_{\rm atm} \xi^4 \frac{\lambda^{(M)}_{{\rm atm}}}{M_\chi^3 M_{\Omega_{\rm atm}}^4}
	+ \overbar{\phi}^{i}_{\rm sol} \overbar{\phi}^{j}_{\rm sol} \xi^5 \frac{\lambda^{(M)}_{{\rm sol}}}{M_\chi^4 M_{\Omega_{\rm sol}}^4}\right]
\end{split}
\label{eq:sYW}
\end{equation}
where $ \lambda_{i,n}^{(f)} $ are constants, $ \overbar{\phi}_{\rm dec} $, $ \overbar{\phi}_{\rm sol} $ and $ \overbar{\phi}_{\rm atm} $ are GUT singlets that are anti-triplets under $ \Delta(27) $, and acquire VEVs according to the CSD3 alignment shown in Eq.~\ref{eq:csd3alignments}. The details of this alignment are discussed in Section \ref{sec:alignment}. The singlet field $ \xi $ acquires a VEV slightly below the GUT scale, and is primarily responsible for the mass hierarchy between fermions through the Froggatt-Nielsen mechanism \cite{Froggatt:1978nt}. 

Each term in the above superpotential has an associated scale derived from the VEVs of the messengers that produce it. These are generally different, but for simplicity we refer to them all as $ M_\chi $ 
when they are produced by pairs of $ SO(10) $ spinor messengers $ \chi $ and $ \overbar{\chi} $.
We make a special note of cases where scale differences have important consequences for the model,
in particular writing $M_{\Omega_{\rm dec}} $, $M_{\Omega_{\rm atm}} $ and $M_{\Omega_{\rm sol}} $
as the combinations of messenger masses that appear in these respective terms. 
This is discussed further in Section \ref{sec:neutrinos} and Fig. \ref{fig:neumass}.

\subsection{Quarks and charged leptons}

\subsubsection{Diagrams and Clebsch-Gordan coefficients}
\label{sec:diagramsandcg}

\begin{figure}[ht]
	\centering
	\begin{subfigure}{0.45\textwidth}
		\centering\includegraphics[scale=0.6]{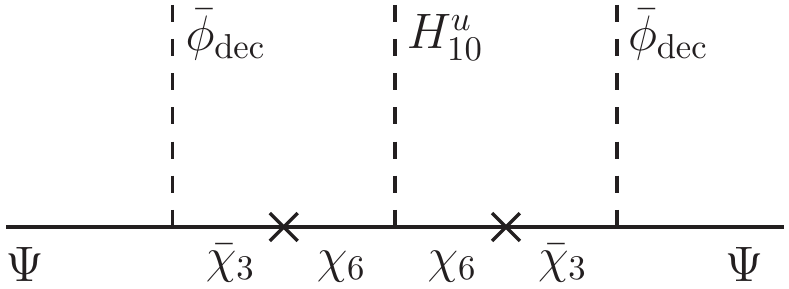}\caption{}
	\end{subfigure}%
	\begin{subfigure}{0.52\textwidth}
		\centering\includegraphics[scale=0.6]{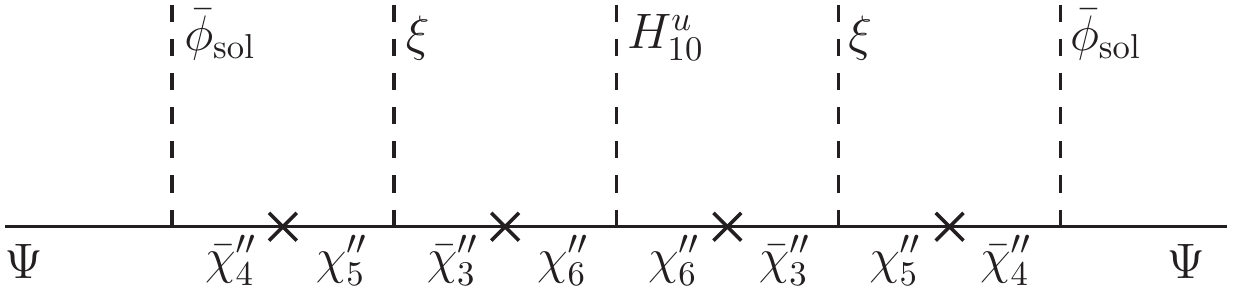}\caption{}
	\end{subfigure}

	\vspace{3ex}
	\begin{subfigure}{0.45\textwidth}
		\centering\includegraphics[scale=0.6]{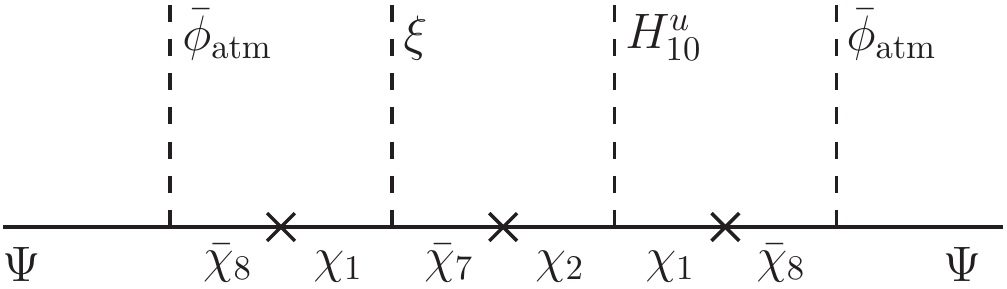}\caption{}
	\end{subfigure}%
	\begin{subfigure}{0.52\textwidth}
		\centering\includegraphics[scale=0.6]{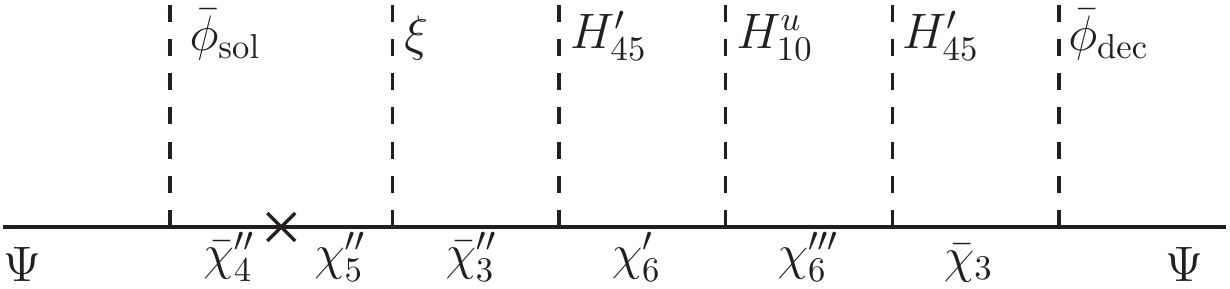}\caption{}
	\end{subfigure}
	\caption{Diagrams for up-type quark and Dirac neutrino Yukawa terms.}
	\label{fig:upmass}
\end{figure}

\begin{figure}[ht]
	\centering
	\begin{subfigure}{0.45\textwidth}
		\centering\includegraphics[scale=0.6]{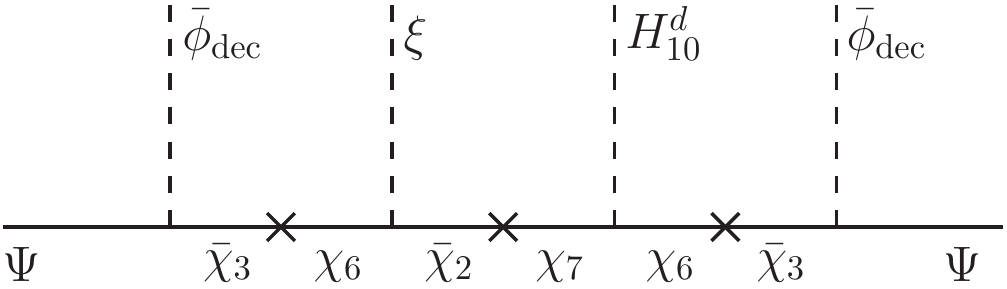}\caption{}
	\end{subfigure}%
	\begin{subfigure}{0.52\textwidth}
		\centering\includegraphics[scale=0.6]{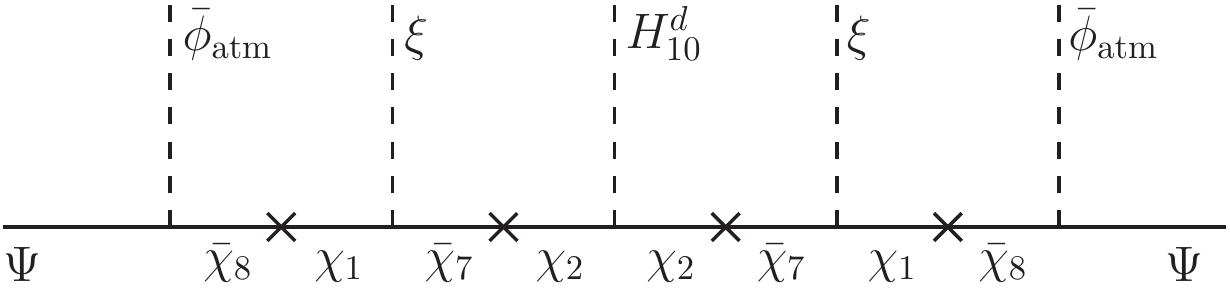}\caption{}
	\end{subfigure}

	\vspace{3ex}
	\begin{subfigure}{\textwidth}
		\centering\includegraphics[scale=0.6]{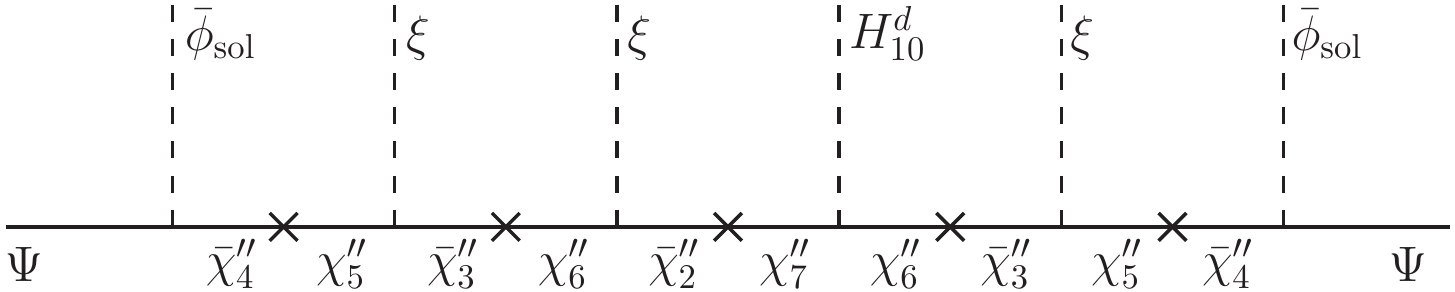}\caption{}
	\end{subfigure}
	\caption{Diagrams for down-type and charged-lepton Yukawa terms.}
	\label{fig:downmass}
\end{figure}

The diagrams involving messengers that give the Yukawa terms in the up sector (first two lines of Eq.~\ref{eq:sYW}) are shown in Fig.~\ref{fig:upmass}, while the diagrams for the down sector (which produce the third and fourth lines of Eq.~\ref{eq:sYW}) are in Fig.~\ref{fig:downmass}. Note that in these and all future diagrams, solid lines correspond to fields with odd $ R $-charge, while dashed lines signify even $ R $-charge.

\begin{figure}[ht]
	\centering
	\includegraphics[scale=0.8]{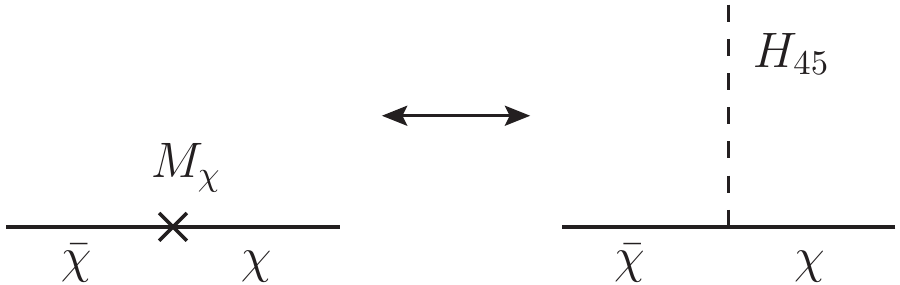}
	\caption{The model symmetries allow for any mass insertion $ M_\chi $ to be replaced by an $ H_{45} \overbar{\chi}\chi $ vertex, leading to extra superpotential terms.}
\label{fig:h45}
\end{figure}

There are several more diagrams that can be written wherein messenger pairs couple to the $ H_{45} $. Specifically,
since the $ H_{45} $ has no $\mathbb{Z}$ charge and is a real representation, it may replace a renormalizable mass diagram as in Fig. \ref{fig:h45}.

The $ H_{45} $ acquiring a VEV leads to Clebsch-Gordan relations, and it will be aligned in such a way that it only affects coloured particles, as will be discussed later.
This is a GUT scale VEV, that we will call $v_{45}$, and breaks $SU(5)\to SU(3)\times SU(2)\times U(1)$.
As an example, consider the charged leptons and down quarks. At the low scale, the superpotential resembles
\begin{equation}
\mathcal{W}_{\rm MSSM} \sim \overbar{d}_L d_R H_{d}\left(\frac{y_1}{M_\chi^2}+\frac{y_2}{v_{45}M_\chi}+\frac{y_3}{v_{45}^2}\right)
+ \overbar{e}_L e_R H_{d} \frac{y_1}{M_\chi^2}.
\label{eq:lincomb}
\end{equation}
We may use the parameters $ y_i $ to fit all the masses.%
\footnote{For the third family we have three $y_i$, with four for the second family, and five for the first family.}
As we take $v_{45}$ to be complex (one of two possibilities, see Section \ref{sec:breaking}), the linear combinations of coefficients $ y_i $ yield a single effective complex coefficient which is typically different for each generation, and different for each of the up, down, charged lepton and neutrino sectors.

\subsubsection{Mass matrices}

\begin{figure}[ht]
	\centering
	\includegraphics[scale=0.8]{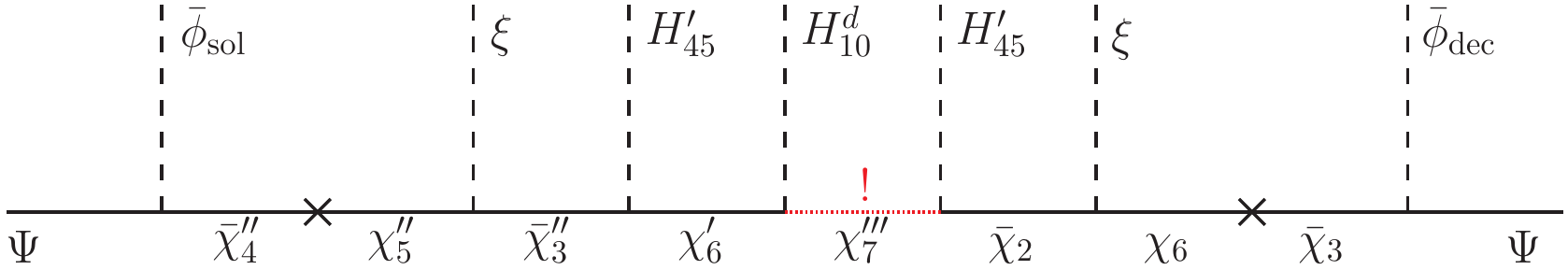}
	\caption{Hypothetical diagram that would produce a mixed term involving $ H^d_{10} $, $ \overbar{\phi}_{\rm sol} $ and $ \overbar{\phi}_{\rm dec} $. Due to an absence of the field $ \chi_{7}^{\prime\prime\prime} $, this term is forbidden.}
	\label{fig:nohd}
\end{figure}

As a consquence of $ SO(10) $ unifying the quarks and leptons, all fermion Dirac matrices have the same generic structure. After the flavons acquire VEVs in the CSD3 alignment, the mass matrices are given by
\begin{equation}
\begin{split}
	m^f
	&= \mu^f_a \braket{\overbar{\phi}_{\rm atm}}^i \braket{\overbar{\phi}_{\rm atm}}^j
		+ \mu^f_s \braket{\overbar{\phi}_{\rm sol}}^i \braket{\overbar{\phi}_{\rm sol}}^j
		+ \mu^f_d \braket{\overbar{\phi}_{\rm dec}}^i \braket{\overbar{\phi}_{\rm dec}}^j \\ 
	&= m^f_a e^{2i \rho_{\rm atm}} \pmatr{0 & 0 & 0 \\ 0 & 1 & 1 \\ 0 & 1 & 1} 
		+ m^f_s e^{2i \rho_{\rm sol}} \pmatr{1 & 3 & 1 \\ 3 & 9 & 3 \\ 1 & 3 & 1}
		+ m^f_d e^{2i \rho_{\rm dec}} \pmatr{0 & 0 & 0 \\ 0 & 0 & 0 \\ 0 & 0 & 1},
\end{split}
\label{eq:massmatrixstructure}
\end{equation}
where $ \mu^f_i $ are coefficients derived from the $ H^{u,d}_{10} $, $ H_{45} $ and $ \xi $ VEVs, and $ \rho_i $ are the phases of flavon VEVs. 
This structure does not include an additional contribution to up quark mass matrix, which arises from a term in $ \mathcal{W}_Y $ (Eq.~\ref{eq:sYW}, line 2). Allowed by the symmetries and messengers, it is proportional to $ \overbar{\phi}_{\rm sol} \overbar{\phi}_{\rm dec} $, and couples to $ H^u_{10} $ but not $ H^d_{10} $. This term leads to the additional contribution to the up quark mass matrix
\begin{equation}
	m^u_{sd}\hspace{1pt}e^{i \rho_{sd}} \pmatr{0&0&1\\0&0&3\\1&3&2} .
\label{eq:musd}
\end{equation}
This mixed term is not allowed for the $H_{10}^d$ due to a lack of messengers able to produce it. In Fig.~\ref{fig:nohd}, we see how this mixed term would have had to be built with an $H_{10}^d$. Since there is no field $\chi_{7}'''$ to build this diagram, it isn't allowed. There are no messengers that allow us to build other mixed terms (involving different pairs of flavons); even if there were, they would be highly suppressed.
Without the term in Eq.~\ref{eq:musd}, the fit to CKM parameters is quite poor, whereas with this term included, a reasonable fit can be made (for more see Section \ref{sec:fit}).

\begin{figure}[ht]
	\centering
	\includegraphics[scale=1]{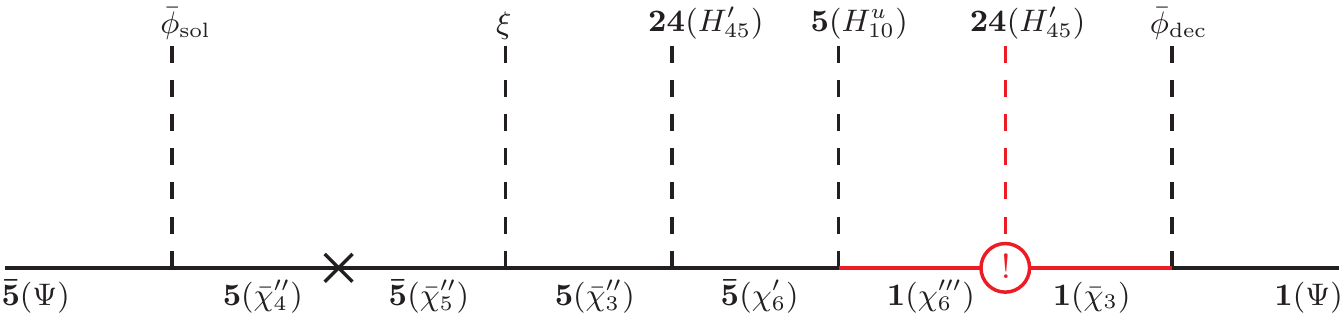}
	\caption{Null contribution from the $ \overbar{\phi}_{\rm sol} \overbar{\phi}_{\rm dec} $ mixed term to neutrinos.}
	\label{fig:crco}
\end{figure}

The additional term in Eq.~\ref{eq:sYW} does not contribute to down quarks or charged leptons, since it only involves $H_{10}^u$. Furthermore, due to its structure it does not contribute to neutrino masses either. 
To see this we may decompose the contribution to neutrinos from the 4th diagram in Fig.~\ref{fig:upmass} in $SU(5)$ terms. 
We adopt the naming convention where the $SU(5)$ representation is labelled by its dimension, with its parent $SO(10)$ field given in parentheses. 
The left handed neutrinos are in $\bar{\textbf{5}}(\Psi)$ and the right handed neutrinos are the $\textbf{1}(\Psi)$.
The diagram would  be in Fig. \ref{fig:crco}. We see that the subdiagram that is emphasized involves one adjoint and two $SU(5)$ singlets, which is zero, therefore the whole diagram is zero.

\subsubsection{Relative phases of flavons}

Since $ H_{45} $ acquires a VEV that only affect coloured particles, the lepton and neutrino Dirac matrices will not depend on $ v_{45} $ (which is generally complex). As such, the only phases contributing to these matrices are $ \rho_{\rm atm} $, $ \rho_{\rm sol} $ and $ \rho_{\rm dec} $, the phases of $ \braket{\overbar{\phi}_{\rm atm}} $, $ \braket{\overbar{\phi}_{\rm sol}} $ and $ \braket{\overbar{\phi}_{\rm dec}} $ respectively, as well as $ \rho_{\xi} $, the phase of $ \braket{\xi} $.
We define the dominant phase as the phase of the subdominant (second) matrix in the seesaw basis where the dominant matrix is real, i.e.
\begin{equation}
	\eta \equiv - \arg\left[\frac{\braket{\overbar{\phi}_{\rm sol}}^2}{\braket{\overbar{\phi}_{\rm atm}}^2} \braket{\xi}\right] = - 2(\rho_{\rm sol} - \rho_{\rm atm}) - \rho_\xi.
	\label{eq:etadef}
\end{equation}
Similarly the subdominant phase is
\begin{equation}
	\eta^\prime \equiv - \arg\left[\frac{\braket{\overbar{\phi}_{\rm dec}}^2}{\braket{\overbar{\phi}_{\rm atm}}^2} \frac{1}{\braket{\xi}}\right] = - 2(\rho_{\rm dec} - \rho_{\rm atm}) + \rho_\xi.
	\label{eq:etaprimedef}
\end{equation}
Each mass matrix derived from the superpotential will have an overall phase dependent on the (generally different) phases of the Higgs VEVs, but these are not physical and may be factored out.
We defined the phases in this way because, as we will see shortly, these definitions are the ones that apply for the effective neutrino mass matrix after seesaw.

This phase structure does not exist in the quark mass matrices, as the factor in front of each submatrix is given as a linear combination of superpotential couplings (see Eq.~\ref{eq:lincomb}), which in turn depend on $v_{45}$. As such, the relative phases in the quark sector are arbitrary.

\subsection{Neutrino masses}
\label{sec:neutrinos}

\begin{figure}[ht]
	\centering
	\begin{subfigure}{\textwidth}
		\centering\includegraphics[scale=0.58]{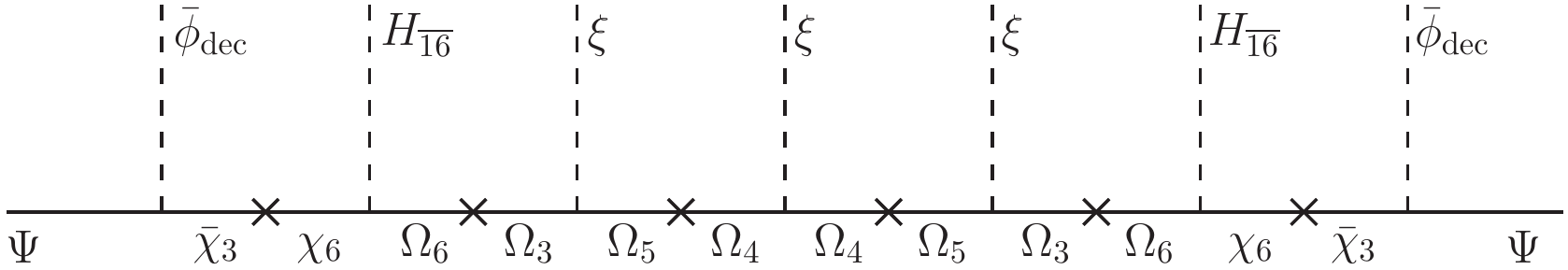}\caption{}
	\end{subfigure}

	\vspace{2ex}
	\begin{subfigure}{\textwidth}
		\centering\includegraphics[scale=0.72]{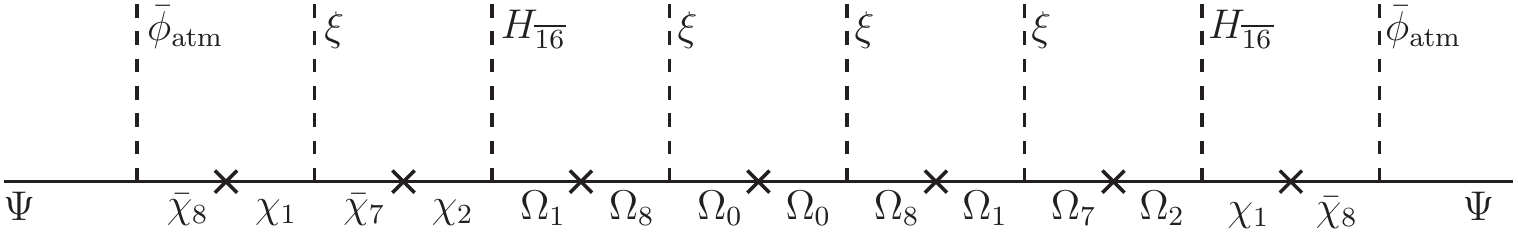}\caption{}
	\end{subfigure}

	\vspace{2ex}
	\begin{subfigure}{\textwidth}
		\centering\includegraphics[scale=0.8]{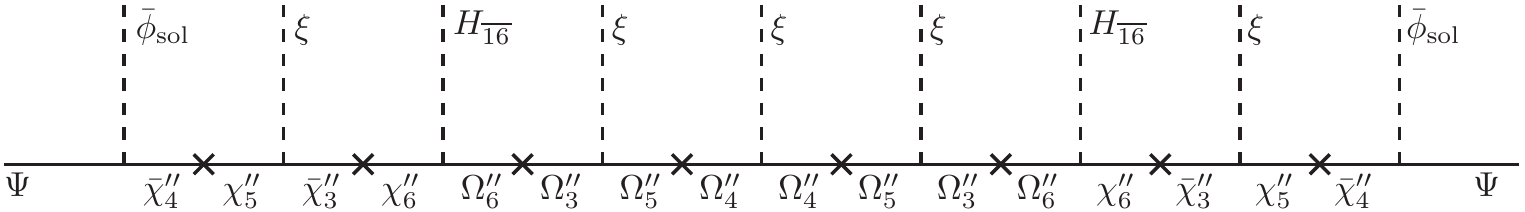}\caption{}
	\end{subfigure}
	\caption{Diagrams for right-handed neutrino Majorana masses.}
	\label{fig:neumass}
\end{figure}

Finally the right-handed neutrino Majorana terms (last line of Eq.~\ref{eq:sYW}) are produced by the diagrams in Fig.~\ref{fig:neumass}. If we decompose these diagrams into $SU(5)$ components, the base line would be all singlets. Therefore there can be no contribution coming from the $H_{45}$ nor the $H_{45}'$ and there is no mixed term allowed. 

Even though they seem quite suppressed, these terms get the correct order. 
It is usual for the right-handed neutrino masses to be in the range $10^{10}-10^{14}$ GeV. 
The VEV $H_{\overbar{16}}$ breaks $SO(10) \to SU(5)$ and thus is higher than the GUT scale, while the scale $ M $ for the messengers is yet higher, such that they may be integrated out. Thus we have $\xi < \braket{H_{\overbar{16}}} \simlt M$ and this way we may obtain the correct scale for right-handed neutrino masses.

It is true, though not immediately obvious, that the mass matrix structure given in Eq.~\ref{eq:massmatrixstructure} is true also for the effective neutrino masses after seesaw.
To show this, consider the neutrino sector after $ SO(10) \rightarrow SU(5) $ breaking, where the left- and right-handed neutrinos $ \nu $ and $ \nu^c $ are contained respectively in a $ \mathbf{\overbar{5}} $ and $ \mathbf{1} $ of $ SU(5) $, in triplets of the family symmetry. We denote the $ \mathbf{\overbar{5}} $ by $ F $ and the singlet by $ N^c $. 
The Dirac mass matrix is then sourced by the terms
\begin{equation}
	H^u_{10} \left[ \frac{\lambda^{(\nu)}_{\rm atm} \xi}{M_\chi^3} (\overbar{\phi}_{\rm atm} F)(\overbar{\phi}_{\rm atm} N^c) + \frac{\lambda^{(\nu)}_{\rm sol} \xi^2}{M_\chi^4} (\overbar{\phi}_{\rm sol} F)(\overbar{\phi}_{\rm sol} N^c) + \frac{\lambda^{(\nu)}_{\rm dec}}{M_\chi^2} (\overbar{\phi}_{\rm dec} F)(\overbar{\phi}_{\rm dec} N^c) \right],
\label{eq:neutrinodirac}
\end{equation}
when the $ H^u_{10} $, $ \xi $ and $ \overbar{\phi} $ fields acquire VEVs. 
Pairs of terms in parentheses, like $ (\phi F) $ and $ (\phi N^c) $, signify a contraction of a $ \Delta(27) $ triplet-antitriplet pair, yielding a flavour singlet. 
In a similar fashion, the right-handed Majorana matrix originates from the terms
\begingroup
\begin{equation}
\begin{split}
	\braket{H_{\overbar{16}} H_{\overbar{16}}}
	\left[
	\frac{\lambda^{(M)}_{\rm atm} \xi^4}{M_\chi^3 M_{\Omega_{\rm atm}}^4} (\overbar{\phi}_{\rm atm} N^c) (\overbar{\phi}_{\rm atm} N^c) \right.
	&+ \frac{\lambda^{(M)}_{\rm sol} \xi^5}{M_\chi^4 M_{\Omega_{\rm sol}}^4} (\overbar{\phi}_{\rm sol} N^c) (\overbar{\phi}_{\rm sol} N^c)
	\\
	&+ \left.\frac{\lambda^{(M)}_{\rm dec} \xi^3}{M_\chi^2 M_{\Omega_{\rm dec}}^4} (\overbar{\phi}_{\rm dec} N^c) (\overbar{\phi}_{\rm dec} N^c) 
	\right],
\end{split}
\label{eq:neutrinomajorana}
\end{equation}
\endgroup
where we have made a distinction between the average scales of the messengers that produce each of the above three terms, giving us three distinct mass scales for the $ \Omega $-type messengers, denoted $ M_{\Omega_{\rm atm}} $, $ M_{\Omega_{\rm sol}} $ and $ M_{\Omega_{\rm dec}} $. We will see that the best fit to data suggests that the third effective neutrino mass is small.
Implementing the seesaw mechanism, requiring the third right-handed neutrino to be decoupled,
means that the last term in Eq.~\ref{eq:neutrinomajorana}, leading to a very large third right-handed neutrino mass  $M_{\rm dec}$, 
can be achieved if $ M_{\Omega_{\rm dec}} < M_{\Omega_{\rm atm}},M_{\Omega_{\rm sol}} $.

We now demonstrate how the seesaw mechanism is implemented in our model.%
\footnote{This is a variation of the mechanism described in \cite{deMedeirosVarzielas:2008en}.}

Collecting the Higgs and $ \xi $ fields along with $ \lambda $ coefficients into generic parameters $ \kappa $ (with dimensions of inverse mass), we can write Eqs.~\ref{eq:neutrinodirac}-\ref{eq:neutrinomajorana} in the simplified form
\begin{equation}
\begin{split}
		& \kappa^{\nu}_{\rm atm} (\overbar{\phi}_{\rm atm} F)(\overbar{\phi}_{\rm atm} N^c) + \kappa^{\nu}_{\rm sol} (\overbar{\phi}_{\rm sol} F)( \overbar{\phi}_{\rm sol}N^c) + \kappa^{\nu}_{\rm dec} ( \overbar{\phi}_{\rm dec}F)( \overbar{\phi}_{\rm dec}N^c) \\
		& \quad + \kappa^{M}_{\rm atm} ( \overbar{\phi}_{\rm atm}N^c)( \overbar{\phi}_{\rm atm}N^c) + \kappa^{M}_{\rm sol} ( \overbar{\phi}_{\rm sol}N^c)( \overbar{\phi}_{\rm sol}N^c)+ \kappa^{M}_{\rm dec} ( \overbar{\phi}_{\rm dec}N^c)( \overbar{\phi}_{\rm dec}N^c),
\end{split}
\label{eq:lambdanu}
\end{equation}
noting also that generically $ \kappa^\nu \ll \kappa^M $. This can be written in matrix form as
\begin{equation}
\begin{blockarray}{c ccc ccc}
	& (\overbar{\phi}_{\rm atm} F) & (\overbar{\phi}_{\rm sol} F) & (\overbar{\phi}_{\rm dec} F) & (\overbar{\phi}_{\rm atm} N^c) & (\overbar{\phi}_{\rm sol} N^c) & (\overbar{\phi}_{\rm dec} N^c) \\[1ex]
	\begin{block}{c(cccccc)}
		(\overbar{\phi}_{\rm atm} F) & 0 & 0 & 0 & \kappa^{\nu}_{\rm atm} & 0 & 0 \\
		(\overbar{\phi}_{\rm sol} F) & 0 & 0 & 0 & 0 & \kappa^{\nu}_{\rm sol} & 0 \\
		(\overbar{\phi}_{\rm dec} F) & 0 & 0 & 0 & 0 & 0 & \kappa^{\nu}_{\rm dec} \\
		(\overbar{\phi}_{\rm atm} N^c) & \kappa^{\nu}_{\rm atm} & 0 & 0 & \kappa^{M}_{\rm atm}  & 0 & 0 \\
		(\overbar{\phi}_{\rm sol} N^c) & 0 & \kappa^{\nu}_{\rm sol} & 0 & 0 & \kappa^{M}_{\rm sol}  & 0 \\
		(\overbar{\phi}_{\rm dec} N^c) & 0 & 0 & \kappa^{\nu}_{\rm dec} & 0 & 0 & \kappa^{M}_{\rm dec}  \\
	\end{block}
\end{blockarray}\,.
\end{equation}
Diagonalisation gives,
to $ \mathcal{O}\!\left((\kappa^{\nu}/\kappa^{M})^{2}\right) $,
the effective Majorana mass terms
\begin{equation}
	-\dfrac{(\kappa^{\nu}_{\rm atm})^2}{\kappa^{M}_{\rm atm}} (\overbar{\phi}_{\rm atm} F)(\overbar{\phi}_{\rm atm} F)
	-\dfrac{(\kappa^{\nu}_{\rm sol})^2}{\kappa^{M}_{\rm sol}} (\overbar{\phi}_{\rm sol} F)(\overbar{\phi}_{\rm sol} F)
	-\dfrac{(\kappa^{\nu}_{\rm dec})^2}{\kappa^{M}_{\rm dec}} (\overbar{\phi}_{\rm dec} F)(\overbar{\phi}_{\rm dec} F).
	\label{eq:lambdanu2}
\end{equation}
These in turn reproduce a light neutrino Majorana mass matrix of the form given in Eq.~\ref{eq:massmatrixstructure}, when the flavons acquire the CSD3 VEVs. 

One final step that is particularly relevant to determining the physical phases is to change to the seesaw basis, as in \cite{Bjorkeroth:2015ora}.
From the neutrino superpotential terms Eqs.~\ref{eq:neutrinodirac}-\ref{eq:neutrinomajorana} we define the neutrino Yukawa and right-handed Majorana matrices $ \lambda^\nu $ and $ M^c $
\begin{equation}
	W_\nu = \lambda^\nu H^u_{10} F N^c + M^c N^c N^c ,
\end{equation}
with the structure of $\lambda^\nu$ and $M^c$ arising directly from the flavon VEVs, as in Eq.~\ref{eq:massmatrixstructure}. This is the SUSY basis.
In the so-called seesaw basis, in a Left-Right (LR) convention, the Yukawa and Majorana matrices $Y^{\nu}$ and $ M_R $ are instead defined by the Lagrangian
\begin{equation}
	\mathcal{L}^{LR} = - H^u_{10} Y^{\nu}_{ij} \overline{L}^i_L  \nu^{j}_R -\tfrac{1}{2} M_{R} \overline{\nu^c}_{R} \nu_{R}+ \mathrm{h.c.},
\end{equation}
where the three families are labeled by $i,j=1,2,3$, $L_i$ are the lepton doublets and $\nu_R^j$ are the right-handed neutrinos below the GUT scale. The light effective Majorana neutrino mass matrix $m^\nu$, defined by
\begin{equation}
\mathcal{L}^{LL}_\nu = -\tfrac{1}{2} m^\nu \overline{\nu}_{L} \nu^{c}_L + \mathrm{h.c.},
\end{equation}
is then determined by the seesaw mechanism
\begin{equation}
	m^{\nu} = v_u^2 Y^{\nu} M^{-1}_{R} Y^{\nu \mathrm{T}}.
\label{eq:seesaw}
\end{equation}
The matrices in the seesaw basis are obtained by complex conjugation of the matrices in the SUSY basis, i.e.
\begin{equation}
Y^{\nu}=(\lambda^{\nu})^* , \qquad M_R=(M^c)^*.
\end{equation} 
We proceed in the seesaw basis, wherein
\begin{equation}
\begin{split}
	Y^\nu
	&= \kappa^{\nu*}_{\rm atm} v_{\rm atm}^{*2} \pmatr{0&0&0\\0&1&1\\0&1&1} 
		+ \kappa^{\nu*}_{\rm sol} v_{\rm sol}^{*2} \pmatr{1&3&1\\3&9&3\\1&3&1}
		+ \kappa^{\nu*}_{\rm dec} v_{\rm dec}^{*2} \pmatr{0&0&0\\0&0&0\\0&0&1},\\
	M_R
	&= \kappa^{M*}_{\rm atm} v_{\rm atm}^{*2} \pmatr{0&0&0\\0&1&1\\0&1&1} 
		+ \kappa^{M*}_{\rm sol} v_{\rm sol}^{*2} \pmatr{1&3&1\\3&9&3\\1&3&1}
		+ \kappa^{M*}_{\rm dec} v_{\rm dec}^{*2} \pmatr{0&0&0\\0&0&0\\0&0&1},	
\end{split}
\end{equation}
using the effective parameters introduced in Eq. \ref{eq:lambdanu}.

To verify that the relative phases are again $ \eta $ and $ \eta^\prime $ (as defined in Eqs.~\ref{eq:etadef} and \ref{eq:etaprimedef}), we may insert VEVs of all fields (denoted $ v_f $ for given field $ f $) to give
\begin{equation}
\medmuskip=3mu
\thinmuskip=0mu
\thickmuskip=0mu
\begin{split}
	m^\nu &~=~ 
	\frac{(v_{H^u_{10}}^*)^2}{(v_{H_{\overbar{16}}}^*)^2}
	\left[
	\frac{(\lambda^{(\nu)}_{\rm atm})^2 M_{\Omega_{\rm atm}}^4}{\lambda^{(M)}_{\rm atm} M_\chi^3}
	\frac{(v_{\rm atm}^{*2} v_\xi^*)^2}{v_{\rm atm}^{*2} v_\xi^{*4}} 
	\pmatr{0 & 0 & 0 \\ 0 & 1 & 1 \\ 0 & 1 & 1} 
	+
	\frac{(\lambda^{(\nu)}_{\rm sol})^2 M_{\Omega_{\rm sol}}^4}{\lambda^{(M)}_{\rm sol} M_\chi^4}
	\frac{(v_{\rm sol}^{*2} v_\xi^{*2})^2}{v_{\rm sol}^{*2} v_\xi^{*5}} 
	\pmatr{1 & 3 & 1 \\ 3 & 9 & 3 \\ 1 & 3 & 1} \right. \\
	&\hspace{13ex}+
	\left.
	\frac{(\lambda^{(\nu)}_{\rm dec})^2 M_{\Omega_{\rm dec}}^4}{\lambda^{(M)}_{\rm dec} M_\chi^2}
	\frac{v_{\rm dec}^{*4}}{v_{\rm dec}^{*2} v_\xi^{*3}} 
	\pmatr{0 & 0 & 0 \\ 0 & 0 & 0 \\ 0 & 0 & 1} \right] \\
	\therefore ~ m^\nu&~\equiv~ \mu_a e^{i \alpha} \pmatr{0 & 0 & 0 \\ 0 & 1 & 1 \\ 0 & 1 & 1}
		+ \mu_b e^{i\beta} \pmatr{1 & 3 & 1 \\ 3 & 9 & 3 \\ 1 & 3 & 1}
		+ \mu_c e^{i\gamma} \pmatr{0 & 0 & 0 \\ 0 & 0 & 0 \\ 0 & 0 & 1},
\end{split}
\end{equation}
where the conjugation of the VEVs is due to changing to the seesaw basis (described above) and we define
\begin{alignat}{3}
	\mu_a &\equiv \left| \frac{(v_{H^u_{10}})^2}{(v_{H_{\overbar{16}}})^2} \frac{(\lambda^{(\nu)}_{\rm atm})^2 M_{\Omega_{\rm atm}}^4}{\lambda^{(M)}_{\rm atm} M_\chi^3} \frac{(v_{\rm atm}^2 v_\xi)^2}{v_{\rm atm}^2 v_\xi^4} \right| ,\qquad && \alpha \equiv - \arg\left[\frac{(v_{H^u_{10}})^2}{(v_{H_{\overbar{16}}})^2} \frac{v_{\rm atm}^2}{v_\xi^2}\right], \nonumber \\
	\mu_b &\equiv \left|  \frac{(v_{H^u_{10}})^2}{(v_{H_{\overbar{16}}})^2} \frac{(\lambda^{(\nu)}_{\rm sol})^2 M_{\Omega_{\rm sol}}^4}{\lambda^{(M)}_{\rm sol} M_\chi^4}
	\frac{(v_{\rm sol}^2 v_\xi^2)^2}{v_{\rm sol}^2 v_\xi^5} \right| , \qquad && \beta \equiv - \arg\left[\frac{(v_{H^u_{10}})^2}{(v_{H_{\overbar{16}}})^2} \frac{v_{\rm sol}^2}{v_\xi}\right], \\
	\mu_c &\equiv  \left| \frac{(v_{H^u_{10}})^2}{(v_{H_{\overbar{16}}})^2} \frac{(\lambda^{(\nu)}_{\rm dec})^2 M_{\Omega_{\rm dec}}^4}{\lambda^{(M)}_{\rm dec} M_\chi^2}
	\frac{v_{\rm dec}^4}{v_{\rm dec}^2 v_\xi^3} \right| , \qquad &&\gamma \equiv - \arg\left[\frac{(v_{H^u_{10}})^2}{(v_{H_{\overbar{16}}})^2} \frac{v_{\rm dec}^2}{v_\xi^3}\right],\nonumber 
\end{alignat}
where messenger masses and $\lambda$ couplings are all real due to CP conservation.
The remarkable fact that the effective left-handed Majorana neutrino mass matrix has the same structure as the neutrino Yukawa matrix and heavy right-handed Majorana neutrino mass matrix can be understood from the argument presented in Eqs.~\ref{eq:lambdanu}-\ref{eq:lambdanu2}.

As before, the physical phases $ \eta $ and $ \eta^\prime $ are defined as the relative phases between the dominant (atm) matrix and, respectively, the subdominant (sol) matrix and sub-subdominant (dec) matrix, i.e.
\begin{equation}
\begin{split}
	\eta &\equiv \beta - \alpha = -\arg\left[\frac{v_{\rm sol}^2}{v_\xi}\right] + \arg\left[\frac{v_{\rm atm}^2}{v_\xi^2}\right] = - 2( \rho_{\rm sol} - \rho_{\rm atm}) - \rho_\xi,\\
	\eta^\prime &\equiv \gamma - \alpha =  - \arg\left[\frac{v_{\rm dec}^2}{v_\xi^3}\right] + \arg\left[\frac{v_{\rm atm}^2}{v_\xi^2}\right] = - 2( \rho_{\rm dec} - \rho_{\rm atm}) + \rho_\xi,
\label{eq:etadefagain}
\end{split}
\end{equation}
which is identical to Eqs.~\ref{eq:etadef}-\ref{eq:etaprimedef}.

\subsection{Renormalisability of the top}
\label{sec:topmass}
The terms in the superpotential in Eq.~\ref{eq:sYW} that are primarily reponsible for the masses of the third family of fermions are, naively,
\begin{equation}
	\Psi_i \Psi_j H_{10}^u 
	\overbar{\phi}^{i}_{\rm dec}\overbar{\phi}^{j}_{\rm dec} 
	\sum_{n=0}^2 \frac{\lambda^{(u)}_{{\rm dec},n}}{\braket{H_{45}}^n M_\chi^{2-n}}.
\end{equation}
When $ \overbar{\phi}_{\rm dec} $ gets a VEV like $ (0,0,v_{\rm dec}) $, with $ v_{\rm dec} $ assumed to be near the GUT scale, these terms reduce to
\begin{equation}
	v_{\rm dec}^2 \Psi_3 \Psi_3 H_{10}^u
	\sum_{n=0}^2 \frac{\lambda^{(u)}_{{\rm dec},n}}{\braket{H_{45}}^n M_\chi^{2-n}}.
\end{equation}

In fact we can only consistently write these non-renormalisable terms when 
$\braket{\overbar{\phi}_{\rm dec}} \ll M_\chi$, but as we will justify in Section \ref{sec:driving}, we actually have 
$\braket{\overbar{\phi}_{\rm dec}} \approx M_\chi$ so the simple integrating out of the messengers is not possible. We actually need to work out the mixing between the messengers and the field $\Psi$. This is only necessary for this term since all of the others involve other flavons that have a VEV 
$\braket{\overbar{\phi}_{\rm atm,sol}}\ll \braket{\overbar{\phi}_{\rm dec}}\approx M_\chi$ and powers of
$\braket{\xi}\ll M_\chi$ that allow a consistent integrating out of the messengers.

To prove that this in fact gives us a renormalisable top mass, it is sufficient to examine the first term in the above sum (with $ n = 0 $). It is sourced by the renormalisable terms
\begin{equation}
	\mathcal{W} \sim \Psi\overbar{\phi}_{\rm dec}\overbar{\chi}_3 
		+ M_\chi \chi_6\overbar{\chi}_3 
		+ H_{10}^u \chi_6\chi_6.
\end{equation}
In matrix form, this gives
\begin{equation}
\mathcal{W} \sim
	\pmatr{\Psi_3 & \chi_6 & \overbar{\chi}_3}
	\pmatr{
		0 & 0 & v_{\rm dec}/2 \\ 
		0 & \braket{H_{10}^u} & M_\chi/2 \\ 
		v_{\rm dec}/2 & M_\chi/2 & 0
	}
	\pmatr{\Psi_3\\ \chi_6 \\ \overbar{\chi}_3}.
\end{equation}
Since $\braket{H_{10}^u} \ll v_{\rm dec} \sim M_{\chi} $, diagonalising this mass matrix reveals two heavy and one light eigenstate, the latter being at the electroweak scale and which we can associate with the third family, and crucially with the top quark. Supposing $v_{\rm dec} \approx M_{\chi} $ (as justified in Section \ref{sec:driving}), the electroweak scale eigenstate is
\begin{equation}
	t \approx \frac{1}{\sqrt{2}} \left( \Psi_3 + \chi_6 \right),
\end{equation}
i.e. the third family up-type fermion, specifically the top quark, is a linear combination of $\Psi_3$ and $\chi_6$, where the latter has a renormalizable coupling to the Higgs. The other eigenstates have a mass at the GUT scale and are therefore identified as messenger eigenstates.

\section{Vacuum alignment in \secheadmath{\Delta(27)}}\label{sec:alignment}

In this section we describe the basic properties of the $ \Delta(27) $ group, and how the CSD3 alignment is produced by $F$-term alignment and orthogonality arguments. We further write down a superpotential which drives the VEVs of the flavons, such that they acquire expectation values at a fixed scale (slightly below the GUT scale), with phases fixed to discrete roots of unity. 
In particular, the relative phases between $ \overbar{\phi}_{\rm atm} $, $ \overbar{\phi}_{\rm sol} $ and $ \overbar{\phi}_{\rm dec} $ are constrained to discrete choices, which subsequently fixes the physical phases $ \eta $, $ \eta^\prime $ in the lepton mass matrices to exact values. 

\subsection{Group products}
The $\Delta(27)$ rules for taking the product of a triplet $ A $ and an antitriplet $ \bar{B} $ are
\begin{equation}
\begin{split}
[A \bar{B}]_{00} &\equiv (a_1 \bar{b}^1 + a_2 \bar{b}^2 + a_3 \bar{b}^3)_{00}\\
[A \bar{B}]_{01} &\equiv (a_1 \bar{b}^3 + a_2 \bar{b}^1 + a_3 \bar{b}^2)_{01}\\
[A \bar{B}]_{02} &\equiv (a_1 \bar{b}^2 + a_2 \bar{b}^3 + a_3 \bar{b}^1)_{02}\\
[A \bar{B}]_{10} &\equiv (a_1 \bar{b}^1 + \omega^2 a_2 \bar{b}^2 + \omega a_3 \bar{b}^3)_{10}\\
[A \bar{B}]_{11} &\equiv (\omega a_1 \bar{b}^3 + a_2 \bar{b}^1 + \omega^2 a_3 \bar{b}^2)_{11}\\
[A \bar{B}]_{12} &\equiv (\omega^2 a_1 \bar{b}^2 + \omega a_2 \bar{b}^3 + a_3 \bar{b}^1)_{12}\\
[A \bar{B}]_{20} &\equiv (a_1 \bar{b}^1 + \omega a_2 \bar{b}^2 + \omega^2 a_3 \bar{b}^3)_{20}\\
[A \bar{B}]_{21} &\equiv (\omega^2 a_1 \bar{b}^3 + a_2 \bar{b}^1 + \omega a_3 \bar{b}^2)_{21}\\
[A \bar{B}]_{22} &\equiv (\omega a_1 \bar{b}^2 + \omega^2 a_2 \bar{b}^3 + a_3 \bar{b}^1)_{22}	
\end{split}
\end{equation}
where $\omega \equiv e^{i 2\pi/3}$.
The product of two triplets or two antitriplets yields, respectively, an antitriplet or a triplet. There are three possible products that can be made in each case, labelled $I$ (identity), $S$ (symmetric) and $A$ (antisymmetric). Defining triplets $ A = (a_1,a_2,a_3) $, $ B = (b_1,b_2,b_3 ) $ and antitriplets $ \bar{A} = (\bar{a}^1,\bar{a}^2,\bar{a}^3) $, $ \bar{B} = (\bar{b}^1,\bar{b}^2,\bar{b}^3 ) $, their products are given by
\begin{equation}
\begin{split}
[A B]_{I} &\equiv (a_1 b_1, a_2 b_2, a_3 b_3)_{02}\\
[\bar{A} \bar{B}]_{I} &\equiv (\bar{a}^1 \bar{b}^1, \bar{a}^2 \bar{b}^2, \bar{a}^3 \bar{b}^3)_{01}\\
[A B]_{S} &\equiv (a_2 b_3 + a_3 b_2, a_3 b_1 + a_1 b_3, a_1 b_2 + a_2 b_1)_{02}\\
[\bar{A} \bar{B}]_{S} &\equiv (\bar{a}^2 \bar{b}^3 + \bar{a}^3 \bar{b}^2, \bar{a}^3 \bar{b}^1 + \bar{a}^1 \bar{b}^3, \bar{a}^1 \bar{b}^2 + \bar{a}^2 \bar{b}^1)_{01}\\
[A B]_{A} &\equiv (a_2 b_3 - a_3 b_2, a_3 b_1 - a_1 b_3, a_1 b_2 - a_2 b_1)_{02}\\
[\bar{A} \bar{B}]_{A} &\equiv (\bar{a}^2 \bar{b}^3 - \bar{a}^3 \bar{b}^2, \bar{a}^3 \bar{b}^1 - \bar{a}^1 \bar{b}^3, \bar{a}^1 \bar{b}^2 - \bar{a}^2 \bar{b}^1) _{01}
\end{split}
\end{equation}
Note that the bar on antitriplet fields serve merely a reminder of their assignment under $\Delta(27)$.

\subsection{CSD3 directions in \secheadmath{\Delta(27)}}

The special directions for $\Delta(27)$ are VEVs with two zeros, and VEVs with 3 equal magnitudes, with phases that are powers of $\omega = e^{i 2\pi/3}$.
There are 3 distinct ways to obtain either the $(0,0,1)$ class of VEV or the $(1,1,1)$ class of VEV  \cite{Varzielas:2015aua}.
One of the possibilities that we make use of here uses invariants built out of an anti-triplet and triplet, and out of three triplets, of the type
\begin{align}
 c [A \overbar{\phi}]_{00} + c_I [A [\phi \phi]_I]_{00} + c_S [A [\phi \phi]_S]_{00}
\end{align}
where $\overbar{\phi}$ is an anti-triplet unrelated with triplet $\phi$ and $A$ is itself a triplet, giving rise to 3 $F$-terms
\begin{equation}
\begin{split}
c \overbar{\phi}^1 + c_I \phi_1 \phi_1 + 2 c_S \phi_2 \phi_3= 0 \\
c \overbar{\phi}^2 + c_I \phi_2 \phi_2 + 2 c_S \phi_3 \phi_1= 0 \\
c \overbar{\phi}^3 + c_I \phi_3 \phi_3 + 2 c_S \phi_1 \phi_2= 0 
\end{split}
\label{Aphiphi}
\end{equation}

To obtain the VEVs we require in the $(0,0,1)$ and $(1,1,1)$ direction class of VEV, an economical solution is the superpotential.
\begin{equation}
\begin{split}
\mathcal{W}_{V0} &=
	c_{a} [\phi_{0} \overbar{A}_0]_{00} \sigma^0_{00} + c_b [\phi_{0} \overbar{A}_0]_{02} \sigma^0_{01} \\
	&\qquad+  c_c [A_{1} \overbar{\phi}_{1}]_{00}M+ c_d [A_{1} \overbar{\phi}_{1}]_{02} \sigma^1_{01} \\
	&\qquad+  c_e [A_3 \overbar{\phi}_3]_{00} M+ c_f [A_3 [\phi_4 \phi_4]_I]_{00} + c_g [A_3 [\phi_4 \phi_4]_S]_{00}  \\
	&\qquad+  c_h [\phi_4 \overbar{A}_4 ]_{00} M + c_i [[\overbar{\phi}_3 \overbar{\phi}_3]_I \overbar{A}_4]_{00} + c_j [[\overbar{\phi}_3 \overbar{\phi}_3]_S \overbar{A}_4]_{00} \\
	&\qquad+ O_{02}[\phi_{2} \overbar{\phi}_{3}]_{01} + O_{00} [\phi_{2} \overbar{\phi}_{1}]_{00}
\label{SV}
\end{split}
\end{equation}
where the $c_{x}$ ($x=a,...,j$) are coefficients that we show explicitly, and the coefficients for the other terms are not shown as they aren't relevant when taking the respective $F$-term.
The triplet flavon $\phi_0$ is aligned to $(1,\omega, \omega^2)$ similarly to how the anti-triplet flavon $\overbar{\phi}_{1}$ is aligned to $(1,1,1)$, through the alignment anti-triplet $\overbar{A}_{0}$ or triplet $A_{1}$ and flavon singlets $\sigma^{0}_{00}$, $\sigma^{0}_{01}$ VEVs with a relative phase of $\omega$ and $\sigma^{1}_{01}$ taking a real VEV.

The anti-triplet flavon $\overbar{\phi}_{3}$ is aligned in a $(0,0,1)$ direction together with triplet $\phi_{4}$. This proceeds from the $F$-terms of the components of $A_3$ and $\overbar{A}_4$, which are of the type shown in Eq.~\ref{Aphiphi}. Taken together the 6 equations only allow a discrete set of solutions where both flavons are aligned in the same direction. One of the solutions has them aligned like $(0,0,v_{3})$ and $(0,0,v_{4})$,%
\footnote{The phenomenologically viable solution is where both flavons are aligned in the $(0,0,1)$ direction, another possibility is that they would both be aligned in the $(1,1,1)$ direction.}
with their magnitudes $v_{3}$ and $v_{4}$ fixed. The relevant VEV magnitudes are
\begin{equation}
	v_{3}^3 = -\frac{c_e c_h^2}{c_f c_i^2}M^3,\qquad
	v_{4}^3 = -\frac{c_e^2 c_h}{c_f^2 c_i}M^3,\qquad
	\braket{\sigma^1_{01}}=-\frac{c_c}{c_d}M.
	\label{eq:vevs34sigma}
\end{equation}
We impose trivial CP symmetry on the fields, including the triplets and anti-triplets. This is consistent with the contractions that make invariants with the $1_{0i}$ set of singlets that we are using. Since the coupling constants $c_x$ are forced to be real by CP conservation, up to minus signs (which can be reabsorbed into the real coefficients) the VEVs $v_{3,4}$ can have a phase only as a third root of unity while $\braket{\sigma^1_{01}}$ has to be real. We expect this mass scale $M$ to be around the GUT scale and with $\mathcal{O}(1)$ $c$ parameters, these VEVs should be at this scale also.

The triplet $\phi_{2}$ is then forced into the $(0,y_2,z_2)$ direction due to the alignment singlet $O_{02}$ and the alignment singlet $O_{00}$ ensures $y_2=-z_2$ by orthogonality with $(1,1,1)$.

\begin{table}
\centering
\footnotesize
\begin{minipage}[b]{0.45\textwidth}
\begin{tabular}{| c | c c | c c c |}
\hline
\multirow{2}{*}{\rule{0pt}{4ex}Field}	& \multicolumn{5}{c |}{Representation} \\
\cline{2-6}
\rule{0pt}{3ex}			& $\Delta(27)$ & $SO(10)$ & $\mathbb{Z}_{9}$ &$\mathbb{Z}_{12}$ & $\mathbb{Z}_4^R$  \\ [0.75ex]
\hline \hline
\rule{0pt}{3ex}%
$\overbar{\phi}_{\rm dec}$ 	& $\bar{3}$ & $1$ & $6$ & $0$ & 0\\
$\overbar{\phi}_{\rm atm}$ 	& $\bar{3}$ & $1$ & $1$ & $0$ & 0\\
$\overbar{\phi}_{\rm sol}$ 	& $\bar{3}$ & $1$ & $5$ & $6$ & 0\\
$\overbar{\phi}_{1}$ 		& $\bar{3}$ & $1$ & $0$ & $4$ & 0\\
$\overbar{\phi}_{7}$ 		& $\bar{3}$ & $1$ & $0$ & $5$ & 0\\
$\phi_{0}$ 			& $3$ & $1$ & $2$ & $6$ & 0\\
$\phi_{2}$ 			& $3$ & $1$ & $3$ & $7$ & 0\\
$\phi_{8}$ 			& $3$ & $1$ & $1$ & $8$ & 0\\
$\phi_{4}$ 			& $3$ & $1$ & $3$ & $0$ & 0\\
$\phi_{6}$ 			& $3$ & $1$ & $0$ & $11$ & 0\\
\rule{0pt}{3ex}%
$\sigma_{00}^0$ 			& $1_{00}$ & $1$ & $0$ & $1$ & 0\\
$\sigma_{01}^0$ 			& $1_{01}$ & $1$ & $0$ & $1$ & 0\\
$\sigma_{01}^1$ 	  	& $1_{01}$ & $1$ & $0$ & $0$ & 0\\[0.5ex]
\hline
\end{tabular}
\caption{Flavon fields.  \label{tab:VEVA1}}
\end{minipage}%
\qquad\begin{minipage}[b]{0.45\textwidth}
\centering
\begin{tabular}{| c | c c | c c c |}
\hline
\multirow{2}{*}{\rule{0pt}{4ex}Field}	& \multicolumn{5}{c |}{Representation} \\
\cline{2-6}
\rule{0pt}{3ex}			& $\Delta(27)$ & $SO(10)$ & $\mathbb{Z}_{9}$ &$\mathbb{Z}_{12}$ & $\mathbb{Z}_4^R$  \\ [0.75ex]
\hline \hline
\rule{0pt}{3ex}%
$A_{1}$ 			& $3$ & $1$ & $0$ & $8$ & 2\\
$A_{3}$ 			& $3$ & $1$ & $3$ & $0$ & 2\\
$\bar{A}_{0}$ 		& $\bar{3}$ & $1$ & $7$ & $5$ & 2\\
$\bar{A}_{4}$ 		& $\bar{3}$ & $1$ & $6$ & $0$ & 2\\
\rule{0pt}{3ex}%
$O_{02}$ 			& $1_{02}$ & $1$ & $0$ & $5$ & 2\\
$O_{00}$ 			& $1_{00}$ & $1$ & $6$ & $1$ & 2\\
$O_{00}^1$ 			& $1_{00}$ & $1$ & $5$ & $5$ & 2\\
$O_{01}^1$ 			& $1_{01}$ & $1$ & $5$ & $0$ & 2\\
$O_{02}^2$ 			& $1_{02}$ & $1$ & $3$ & $1$ & 2\\
$O_{02}'^2 $ 		& $1_{02}$ & $1$ & $6$ & $0$ & 2\\
$O_{00}^2$ 			& $1_{00}$ & $1$ & $0$ & $8$ & 2\\
$O_{01}^2$ 			& $1_{01}$ & $1$ & $0$ & $8$ & 2\\
$O_{00}^3$ 			& $1_{00}$ & $1$ & $8$ & $11$ & 2\\
$O_{00}'^3$ 		& $1_{00}$ & $1$ & $7$ & $4$ & 2\\
$O_{01}^4$ 			& $1_{01}$ & $1$ & $1$ & $11$ & 2\\
$O_{00}^4$ 			& $1_{00}$ & $1$ & $3$ & $10$ & 2\\[0.5ex]
\hline
\end{tabular}
\caption{Alignment field content.\label{tab:VEVA}}
\end{minipage}
\end{table}

In order to have CSD3 we want the directions $(0,1,1)$ and $(1,3,1)$. We can use a chain of orthogonality relations, where in $\Delta(27)$ they must be between triplet and anti-triplets.
Using the 3 directions above we can arrive relatively easily to $(0,1,1)$, through orthogonality with $\phi_{2}$ and $\phi_{4}$
\begin{align}
\mathcal{W}_{V1} &= 
O^1_{00} [\phi_{2} \overbar{\phi}_{5}]_{00} + O^1_{01} [\phi_{4} \overbar{\phi}_{5}]_{02}. \label{SV1}
\end{align}
With this we obtain a $\overbar{\phi}_{5}$ anti-triplet in the $(0,1,1)$ direction (note the $[]_{02}$ contraction matches the first component of the anti-triplet with the third component of the triplet, putting the zero in the right place in $\overbar{\phi}_{5}$).

In order to get to $(1,3,1)$ we require a $(2,-1,1)$ direction, which itself requires $(1,1,-1)$. To obtain the latter we also duplicate the $\overbar{\phi}_{5}$ direction into a triplet $\phi_{6}$ which is a different field, and unrelated to $\overbar{\phi}_{5}$ other than them having VEVs in the same direction.
A way to do both things in one step is
\begin{align}
	\mathcal{W}_{V2} &= O^2_{02} [\phi_{6} \overbar{\phi}_{3}]_{01} + O'^2_{02} [\phi_{2} \overbar{\phi}_{7}]_{01} +  O^2_{00} [\phi_{6} \overbar{\phi}_{7}]_{00} + O^2_{01} [\phi_{6} \overbar{\phi}_{7}]_{02}. \label{SV2}
\end{align}
Starting with the first two orthogonalities we ensure the zero is in a specific component for $\phi_{6}$ as $(0,y_6,z_6)$ and that $\overbar{\phi}_{7}$ is in the $(x_7,x_7,z_7)$ direction. The other two mutual orthogonalities give $0 x_7 + y_6 x_7 + z_6 z_7 = 0$ and $0 x_7 + y_6 z_7 + z_6 x_7 =0$ which completes the $(0,1,1)$ and $(1,1,-1)$ alignments. Strictly speaking this alignment allows both an undesired solution where we get $(0,1,-1)$ with $(1,1,1)$ and the desired solution of $(0,1,1)$ with $(1,1,-1)$.

The  next step is obtaining the $(2,-1,1)$ as a triplet. For this we want to use the $(0,1,1)$ anti-triplet direction, and the anti-triplet with the recently obtained $(1,1,-1)$ direction.
\begin{align}
	\mathcal{W}_{V3} &= O^3_{00} [\phi_{8} \overbar{\phi}_{7}]_{00} + O'^3_{00} [\phi_{8} \overbar{\phi}_{5}]_{00},
\end{align}
Finally, by orthogonality 
\begin{align}
	\mathcal{W}_{V4} &= O^4_{01} [\phi_{2} \overbar{\phi}_{9}]_{02} + O^4_{00} [\phi_{8} \overbar{\phi}_{9}]_{00},
\end{align}
one obtains the $(1,3,1)$ direction as an anti-triplet.
We did not need to align a $(1,0,-1)$ direction as the $[\ldots ]_{02}$ contraction with the triplet $(0,1,-1)$ ($\phi_{2}$) puts its zero together with the second component of the anti-triplet $\overbar{\phi}_{9}$. 

Noting now that the VEVs of anti-triplets $\overbar{\phi}_{3}$, $\overbar{\phi}_{5}$  and $\overbar{\phi}_{9}$  are the desired directions for $\overbar{\phi}_{\rm dec}$, $\overbar{\phi}_{\rm atm}$, and $\overbar{\phi}_{\rm sol}$ respectively, we now rename these fields to match the notation used in other sections, so $v_3=v_{\rm dec}$ and
\begin{align}
	\overbar{\phi}_{3} \equiv \overbar{\phi}_{\rm dec} ,\quad \overbar{\phi}_{5} \equiv \overbar{\phi}_{\rm atm} ,\quad \overbar{\phi}_{9} \equiv \overbar{\phi}_{\rm sol}.
\end{align}
This notation is also used in Table \ref{tab:VEVA}, which summarises the field content and their representation under the symmetries.
For the sake of completeness we collect all alignment terms into one superpotential
\begin{align}
\mathcal{W}_{V}=\mathcal{W}_{V0}+\mathcal{W}_{V1}+\mathcal{W}_{V2}+\mathcal{W}_{V3}+\mathcal{W}_{V4},
\end{align}
such that, omitting the coefficients, we have
\begin{equation}
\begin{split}
	\mathcal{W}_{V} &\sim
	[\phi_{0} \overbar{A}_0]_{00} \sigma^0_{00} + [\phi_{0} \overbar{A}_0]_{02} \sigma^0_{01} 
	+[A_{1} \overbar{\phi}_{1}]_{00} M+ [A_{1} \overbar{\phi}_{1}]_{02} \sigma^1_{01}  \\
	&\qquad+[A_3 \overbar{\phi}_3]_{00}M + [A_3 [\phi_4 \phi_4]_I]_{00} + [A_3 [\phi_4 \phi_4]_S]_{00} \\
	&\qquad+[\phi_4 \overbar{A}_4 ]_{00} M+ [[\overbar{\phi}_3 \overbar{\phi}_3]_I \overbar{A}_4]_{00} + [[\overbar{\phi}_3 \overbar{\phi}_3]_S \overbar{A}_4]_{00} \\
	&\qquad+O_{02}[\phi_{2} \overbar{\phi}_{3}]_{01} + O_{00} [\phi_{2} \overbar{\phi}_{1}]_{00}
	+ O^1_{00} [\phi_{2} \overbar{\phi}_{\rm atm}]_{00} + O^1_{01} [\phi_{4} \overbar{\phi}_{\rm atm}]_{02} \\
	&\qquad + O^2_{02} [\phi_{6} \overbar{\phi}_{\rm dec}]_{01} + O'^2_{02} [\phi_{2} \overbar{\phi}_{7}]_{01} +  O^2_{00} [\phi_{6} \overbar{\phi}_{7}]_{00} + O^2_{01} [\phi_{6} \overbar{\phi}_{7}]_{02} \\
	&\qquad + O^3_{00} [\phi_{8} \overbar{\phi}_{7}]_{00} + O'^3_{00} [\phi_{8} \overbar{\phi}_{\rm atm}]_{00}
	+ O^4_{01} [\phi_{2} \overbar{\phi}_{\rm sol}]_{02} + O^4_{00} [\phi_{8} \overbar{\phi}_{\rm sol}]_{00}.
\end{split}
\end{equation}

We summarise the alignments produced by the above superpotential as follows: 
\addtocounter{equation}{1}
\begin{empheq}{alignat=4}
\braket{\phi}_{0} 
	&\propto (1,\omega,\omega^2), \hspace{8ex}
	&\braket{\overbar{\phi}_{1}} 
	&\propto (1,1,1), \nonumber\\
\braket{\phi_{2}} 
	&\propto (0,1,-1), \quad
	&\braket{\overbar{\phi}_{\rm dec}} 
	&\propto (0,0,1), \nonumber\\
\braket{\phi_{4}} 
	&\propto (0,0,1), \quad
	&\braket{\overbar{\phi}_{\rm atm}} 
	&\propto (0,1,1),\\
\braket{\phi_{6}} 
	&\propto (0,1,1),
	&\braket{\overbar{\phi}_{7}} 
	&\propto (1,1,-1), \nonumber\\
\braket{\phi_{8}} 
	&\propto (2,-1,1),
	&\braket{\overbar{\phi}_{\rm sol}} 
	&\propto (1,3,1).\nonumber
\end{empheq}

\subsection{Driving flavon VEVs and phases \label{sec:driving}}

\begin{table}[ht]
\centering
\footnotesize
\begin{tabular}{| c | c c | c c c |}
\hline
\multirow{2}{*}{\rule{0pt}{4ex}Field}	& \multicolumn{5}{c |}{Representation} \\
\cline{2-6}
\rule{0pt}{3ex}			& $\Delta(27)$ & $SO(10)$ & $\mathbb{Z}_{9}$ &$\mathbb{Z}_{12}$ & $\mathbb{Z}_4^R$ \\ [0.75ex]
\hline \hline
\rule{0pt}{3ex}%
$ P_1 $ &	$1_{00}$ & 1 & 8 & 1 & 2 \\
$ P_2 $ &	$1_{00}$ & 1 & 1 & 6 & 2 \\
$ P_3 $ &	$1_{01}$ & 1 & 2 & 0 & 2 \\
[0.5ex]
$\zeta_i$ &			$1_{00}$ & 1 & $i\in\{0,1,2,3\}$ & 1 & 2 \\
$ \bar{\zeta}_i $ &		$1_{00}$ & 1 & $i\in\{0,6,7,8\}$& 11 & 0 \\
$\zeta_i'$ &			$1_{01}$ & 1 & $i\in\{3,4,5\}$ & 0 & 2 \\
$ \bar{\zeta}_i' $ &		$1_{02}$ & 1 & $i\in\{4,5,6\}$& 0 & 0 \\
[0.5ex]
\hline
\end{tabular}
\caption{Field content for driving the flavon VEVs.}
\label{tab:pfields}
\end{table}

To drive the flavon VEVs, we introduce a set of fields given in Table \ref{tab:pfields}.
They are GUT singlets with nontrivial representations under $ \Delta(27) $ and the $ \mathbb{Z} $ symmetries, and couple to the flavon fields.

To obtain the necessary superpotential we need to add more messengers $\zeta, \bar{\zeta}$, with a characteristic mass $M_\zeta$, also listed in Table \ref{tab:pfields}.
The superpotential which drives the flavons is
\begin{equation}
\begin{split}
	\mathcal{W}_{\rm \phi} 
		&= P_1\left[\kappa_1\left(\frac{\xi}{M_\zeta}\right)^4\bar{\phi}_{\rm dec}\phi_6-\kappa_2\bar{\phi}_{\rm atm}\phi_6\right]+P_2\left[\kappa_3 \bar{\phi}_{\rm sol}\phi_4-\kappa_4\bar{\phi}_{\rm dec}\phi_0\right]
		\\ &\qquad+P_3\left[ \kappa_5 \bar{\phi}_{\rm sol} \phi_0-\kappa_6 \left(\frac{\xi}{M_\zeta}\right)^3 \bar{\phi}_{\rm atm}\phi_4 \right],
		\label{eq:phasefixing}
\end{split}
\end{equation}
where $\kappa_i$ are real dimensionless constants. As discussed in Section \ref{sec:fit}, to acquire a good fit to data without tuning, we need to assume that $ \braket{\xi} \simlt M_\zeta$. The $F$-term equations for the $P$ fields give relationships between the VEVs of the flavons that couple to the SM fields.  
The (nontrivial) representations of the $P$ fields under $\Delta(27)$ are chosen specifically so that the pairs of flavon VEVs they are multiplied by do not give zero when they acquire VEVs. 

The constants $\kappa_i$ are forced to be real by CP conservation but the VEV $\braket{\phi_0}$ has complex components that introduce phases to the other VEVs. Specifically, the terms multiplied by the constants $\kappa_{4,5}$ obtain the following factors when contracting the $ \Delta(27) $ triplets:
\begin{equation}
[\braket{\bar{\phi}_{\rm sol}}\braket{\phi_0}]_{02} = 2 v_{\rm sol} v_0,
\qquad
[\braket{\bar{\phi}_{\rm dec}}\braket{\phi_0}]_{00} = \omega^2 v_{\rm dec} v_0,
\end{equation}
so we may effectively treat as $\kappa_4$ carrying a factor of $\omega^2$.

We proceed to solve the $F$-term equations coming from the $P$ fields, yielding VEVs for the important flavons $ \overbar{\phi}_{\rm sol} $ and $ \overbar{\phi}_{\rm atm} $, while $ \braket{\overbar{\phi}_{\rm dec}} $ is given in Eq.~\ref{eq:vevs34sigma} (recall that $ v_3 \equiv v_{\rm dec} $). It is useful to note the relation 
$ v_{4} = c_j v^2_{\rm dec} / ( c_h M ) $,
which can be seen from comparing the VEVs in Eq.~\ref{eq:vevs34sigma}. We obtain
\begin{equation}\begin{split}
v_{\rm sol}^2
	= \omega^2\frac{\kappa_4\kappa_6c_h}{2 \kappa_3\kappa_5 c_j}\left(\frac{\xi}{M_\zeta}\right)^7 v_{\rm dec}^2,
	\qquad
v_{\rm atm}^2
	=\frac{\kappa^2_1}{4\kappa^2_2}\left(\frac{\xi}{M_\zeta}\right)^8v_{\rm dec}^2,
\end{split}\end{equation}
where, since $\braket{\xi}/M_\zeta < 1$, we conclude that $v_{\rm dec}\gg v_{\rm atm}\sim v_{\rm sol}$. 
Given these VEVs, the physical phases defined in Eqs.~\ref{eq:etadef}, \ref{eq:etaprimedef} are given by
\begin{equation}
\begin{split}
\eta &= 
	- \arg\left[\frac{v_{\rm sol}^2}{v_{\rm atm}^2} \braket{\xi}\right]=-\arg[\omega^2], \\ 
\eta^\prime &=
	- \arg\left[\frac{v_{\rm dec}^2}{v_{\rm atm}^2} \frac{1}{\braket{\xi}}\right] =9\arg[\braket{\xi}],
\end{split}
\end{equation}
where the real coupling constants $c_x,\kappa_i$ do not contribute to phases. These phases are in fact completely fixed. As will be shown in Eq.~\ref{eq:xivev}, the phase of $\braket{\xi}$ is a ninth root of unity; by the cancellation of this phase we finally have
\begin{equation}
\eta=\frac{2\pi}{3},\qquad\eta'=0.
\label{eq:physicalphases}
\end{equation}
Strictly speaking these phases are fixed only up to a relative phase $ \pi $, depending on the signs of the real constants. However, this additional phase is unphysical, as it may always be subsumed into other real parameters at the low scale.

\section{GUT breaking}
\label{sec:gutbreaking}
In this section we detail how the $ SO(10) $ GUT is broken down to the MSSM via $ SU(5) $, show how doublet-triplet splitting is achieved, and how only two light Higgs doublets are present below the GUT scale,
as in the MSSM. 

\subsection{Breaking potential and diagrams \label{sec:breaking}}
\begin{figure}[ht]
	\centering
	\begin{subfigure}{0.5\textwidth}
		\centering\includegraphics[scale=0.7]{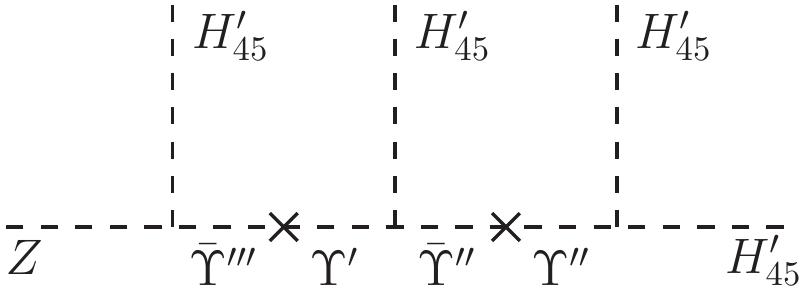}\caption{}
	\end{subfigure}%
        \begin{subfigure}{0.4\textwidth}
		\centering\includegraphics[scale=0.7]{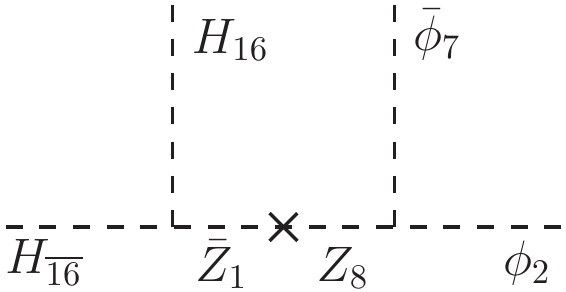}\caption{}
	\end{subfigure}

	\vspace{2ex}
	\begin{subfigure}{\textwidth}
		\centering\includegraphics[scale=0.7]{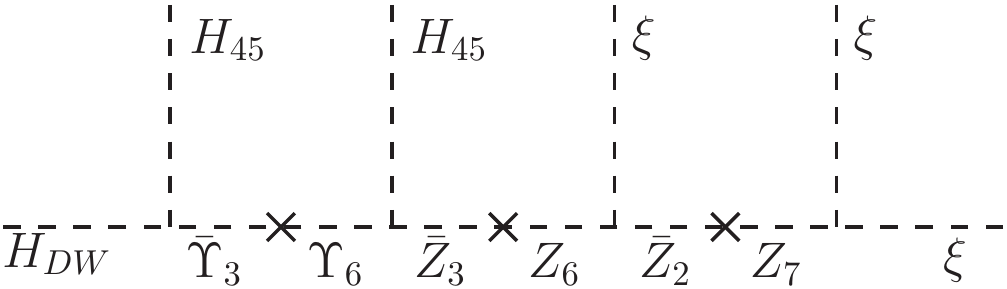}\caption{}
	\end{subfigure}

	\vspace{2ex}
	\begin{subfigure}{\textwidth}
		\centering\includegraphics[scale=0.7]{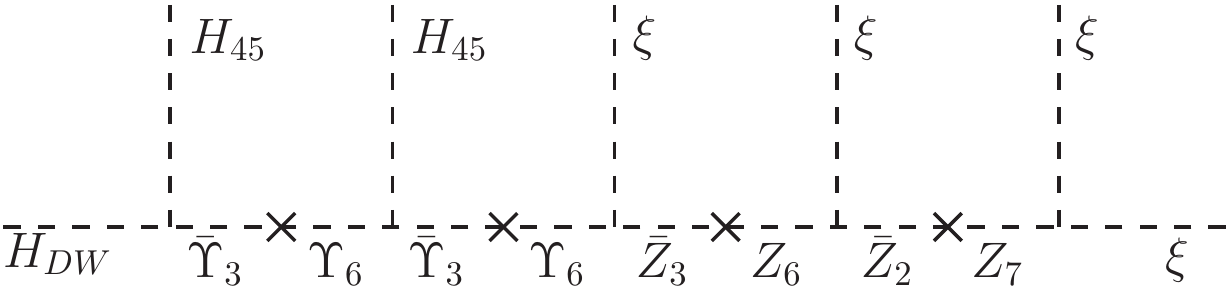}\caption{}
	\end{subfigure}	
	\caption{Diagrams that give rise to GUT breaking terms.}
	\label{fig:nonrenormgutbreaking}
\end{figure}

\begin{figure}[ht]
	\centering
	\begin{subfigure}{\textwidth}
		\centering\includegraphics[scale=0.7]{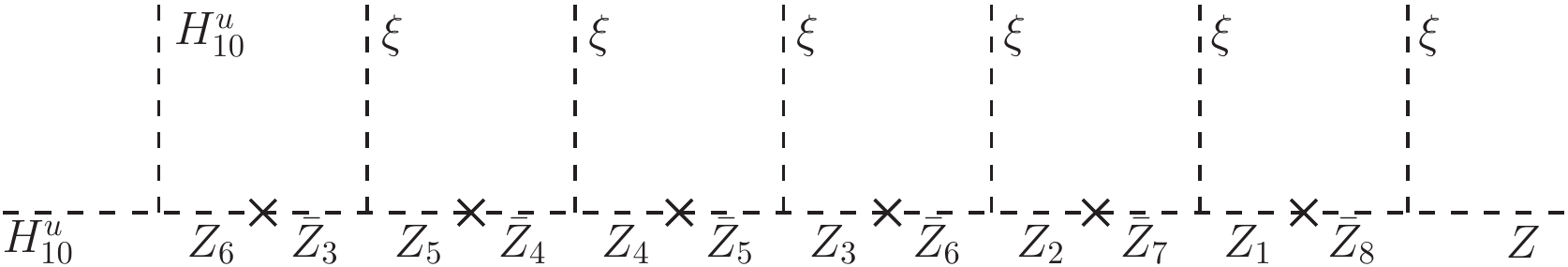}\caption{}
	\end{subfigure}
	
	\vspace{2ex}
	\begin{subfigure}{0.4\textwidth}
		\centering\includegraphics[scale=0.7]{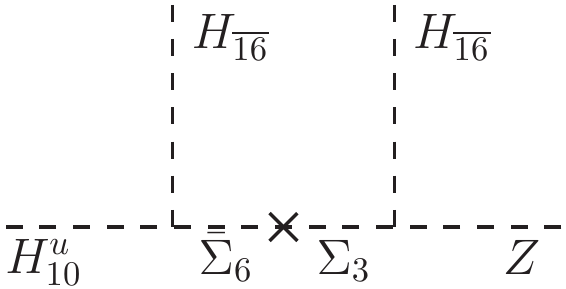}\caption{}
	\end{subfigure}%
	\begin{subfigure}{0.4\textwidth}
		\centering\includegraphics[scale=0.7]{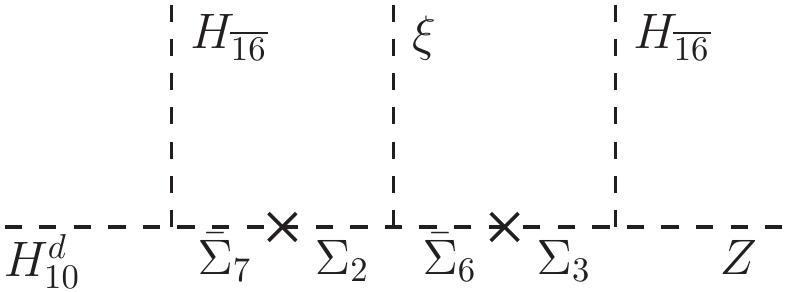}\caption{}
	\end{subfigure}

	\vspace{2ex}
	\begin{subfigure}{0.4\textwidth}
		\centering\includegraphics[scale=0.7]{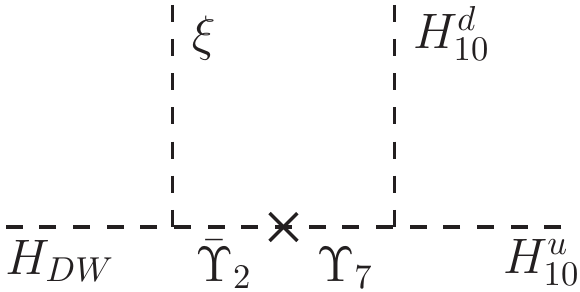}\caption{}
	\end{subfigure}%
	\begin{subfigure}{0.4\textwidth}
		\centering\includegraphics[scale=0.7]{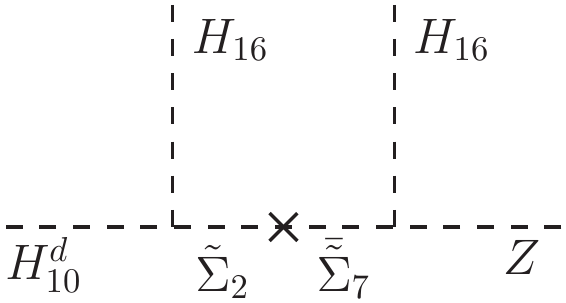}\caption{}
	\end{subfigure}
	\caption{Diagrams that give rise to doublet-triplet splitting.}
	\label{fig:dtsplitting}
\end{figure}

The superpotential that breaks $ SO(10) $ is given by
\begingroup
\medmuskip=2mu
\thinmuskip=2mu
\thickmuskip=2mu
\begin{equation}
\begin{split}
	\mathcal{W}_{\rm GUT}&=
	M^2 Z+\lambda_1 Z^3+\lambda_2 Z Z''^2+\lambda_3Z'' H_{45}'^2+\lambda_4Z\frac{H'^4_{45}}{M_\Upsilon^2}
	\\ &\quad 
	+ \frac{Z}{M_\Sigma} 
	\left(\lambda_5H_{16} H_{16}H^d_{10}+\lambda_6\frac{\xi^8}{M_\Sigma^8}H_{16}H_{16}H^u_{10}+\lambda_7H_{\overbar{16}}H_{\overbar{16}}H^u_{10}+\lambda_8\frac{\xi}{M_\Sigma}H_{\overbar{16}}H_{\overbar{16}}H^d_{10}\right)
	\\ &\quad+\lambda_9 ZH_{DW}^2\frac{\xi^6}{M_Z^6}+H_{DW}\frac{\xi^3}{M_Z^2}\left(\lambda_{10}H_{45}+\lambda_{11}\frac{H_{45}^3}{M_\Upsilon^2}\right)+H_{16} H_{\overbar{16}}\left(\lambda_{12}\xi+\frac{\lambda_{13}}{M_Z}\overbar{\phi}_1\phi_8\right)
	\\ &\quad 
	+ Z\left(\lambda_{14}\frac{\xi^6}{M_Z^6}\overbar{\phi}_7\phi_2+\lambda_{15}\frac{\xi^8}{M_Z^8}\overbar{\phi}_1\phi_8+\lambda_{16}\frac{\xi^5}{M_Z^5}\bar{\phi}_{\rm sol}\phi_4+\lambda_{17}\frac{\xi^2}{M_Z^2}\bar{\phi}_{\rm sol}\phi_0+\lambda_{18}\bar{\phi}_{\rm dec}\phi_4\right).
\end{split}
\label{eq:gutbreak}
\end{equation}
\endgroup
The renormalisable diagrams that give rise to this superpotential are given%
\footnote{We omit those diagrams with seven or eight powers of $ \xi $, as they are constructed in a similar way using the same messengers but are not particularly illuminating.}
in Fig.~\ref{fig:nonrenormgutbreaking} (giving lines 1 and 3) and Fig.~\ref{fig:dtsplitting} (giving line 2), and the corresponding messenger fields ($\Sigma$, $\Upsilon$ and $Z_i$) are detailed in Table \ref{tab:messengers}. Most fields are familiar from the Yukawa sector discussed previously, while the field $ H_{DW} $ is an $ SO(10) $ adjoint that governs doublet-triplet splitting, as we will see shortly. Requiring that every field's $F$-term vanishes yields a set of equations that fixes the VEVs of the above fields.

\begin{table}
\centering
\footnotesize
\begin{tabular}{| c | c c | c c c |}
\hline
\multirow{2}{*}{\rule{0pt}{4ex}Field}	& \multicolumn{5}{c |}{Representation} \\
\cline{2-6}
\rule{0pt}{3ex}			& $\Delta(27)$ & $SO(10)$ & $\mathbb{Z}_{9}$ &$\mathbb{Z}_{12}$ & $\mathbb{Z}_4^R$ \\ [0.75ex]
\hline \hline
\rule{0pt}{3ex}%
$Z_i$ & 1 & 1 & $ i \in \{1,...,8\} $ & 0 &2\\
$\bar{Z}_i$ & 1 & 1 & $ i \in \{1,...,8\} $ & 0 &0\\
\rule{0pt}{3ex}%
$\overbar{\Sigma}_{6,7} $ & 1 & $\overbar{16}$ & $6,7$ & 0 &2\\
$\Sigma_{3,2} $ & 1 & 16 & $3,2$ & 0 &0\\
\rule{0pt}{3ex}%
$\tilde{\Sigma}_i$ & 1 & 16 & $ i \in \{ 0, . . .  ,8\} $ & 0 &2\\
$\overbar{\tilde{\Sigma}}_i$ & 1 & $\overbar{16}$ & $ i \in \{ 0,. . . ,8\} $ & 0 &0\\
\rule{0pt}{3ex}%
$\overbar{\Upsilon}_{3,2}$ & 1 & 45 & $3,2$ & 0 &0\\
$\Upsilon_{6,7} $ & 1 & $45$ & $6,7$ & 0 &2\\
$\overbar{\Upsilon}'''$ & 1 & 45 & $0$ & 9 &0\\
$\Upsilon'$ & 1 & $45$ & $0$ & 3 &2\\
$\overbar{\Upsilon}''$ & 1 & 45 & $0$ & 6 &0\\
$\Upsilon''$ & 1 & $45$ & $0$ & 6 &2\\
\hline
\end{tabular}
\caption{Messenger superfields required for the doublet and triplet mass terms. Note that the model also includes
$Z_0$ (but not $\bar{Z}_0$) which we considered earlier as $Z$, and that $\Upsilon_6$ has the same quantum number as $H_{DW}$ (see Table \ref{tab:funfields}).}
\label{tab:messengers}
\end{table}

The first line contains terms involving different powers of $Z$, $Z''$ and $H_{45}'$, which ensures that their corresponding $F$-term conditions fix all VEVs to be non-zero. The exact expressions for the VEVs are complicated and thus are not shown, since they are not enlightening.

The second line has terms involving the fields $H_{10}^{u,d}$ that will be discussed carefully in the next section on doublet-triplet splitting. At this level, the fields $H_{10}^{u,d}$ have a zero VEV, so any term involving two of them does not contribute to the $F$-term equations. The $F$-term conditions coming from $H_{10}^{u,d}$ themselves relate the $H_{16,\overbar{16}}$ VEVs and also fixes the VEV of $ \xi $ to be
\begin{equation}
	\braket{\xi}=\left(\frac{\lambda_5\lambda_7}{\lambda_8\lambda_6}\right)^{1/9} M_{\Sigma},
\label{eq:xivev}
\end{equation}
which subsequently fixes the phase of $ \braket{\xi} $ to be one of the ninth roots of unity.

At this stage it is relevant to consider superfields $H_{DW}$ and $\Upsilon_6$, which have the same quantum numbers. In terms of superfields $\Upsilon_6^a$, $\Upsilon_6^b$, the mass term for the messenger pair reads $M_\Upsilon (c_a \Upsilon_6^a + c_b \Upsilon_6^b) \overbar{\Upsilon}_3$. We define $\Upsilon_6 \equiv (c_a \Upsilon_6^a + c_b \Upsilon_6^b)$ and $H_{DW}$ as the orthogonal combination. The $F$-term with respect to $\overbar{\Upsilon}_3$ forces $\Upsilon_6$ to have a zero VEV, meaning it won't contribute elsewhere and justifies identifying it as half of the messenger pair.
Therefore, the third line contains different powers of $H_{DW}$ and $H_{45}$ and gives them VEVs. 

The model actually allows an infinity of terms involving $H_{45}$, each with a higher power of this field. We keep only the first two terms since they are enough to give the $H_{45}$ a general VEV, whereas adding the other terms will make its VEV look more complicated, but will not affect the physics. Its own $F$-term equation fixes its VEV to be
\begin{equation}
	v_{45}=\sqrt{-\frac{\lambda_{10}}{\lambda_{11}} } M_{\Upsilon},
\label{eq:h45vev}
\end{equation}
which must define the GUT scale, while we may choose the signs of $\lambda_{10,11}$ so that it is imaginary (this is the phenomenologically favoured solution).
The $F$-term for $\xi$ will fix the VEV of $H_{16,\overbar{16}}$. 
The $F$-terms coming from $H_{16,\overbar{16}}$ will drive the VEVs of the flavons $ \overbar{\phi}_7 $ and $ \phi_2 $ (seen on line 2). 

The last line, allowed by the symmetries and messengers, only adds terms to the $F$-terms for $Z$ and $\xi$, relating their VEVs to the flavon ones. 
The flavon $F$-terms will fix some of the $O$ field VEVs. 

The VEVs $\braket{H_{16,\overbar{16}}}$ specifically break $SO(10)\to SU(5)$. 
The VEVs $\braket{H_{45},H_{45}',H_{DW}}$ specifically break $SU(5)\to SM$. 
The VEV $\braket{\xi}$ completely breaks $\mathbb{Z}_9$. 
Finally, the VEVs $\braket{Z,Z''}$, carrying 2 units of charge under $\mathbb{Z}_4^R$, break it into the usual $\mathbb{Z}_2^R$ R-parity at the GUT scale.

\subsection{Obtaining two light Higgs doublets}
\label{sec:dtsplitting}

In this section we show how the $ SU(2) $ doublets and $ SU(3) $ triplet Higgses contained within the $ H_{16,\overbar{16}} $ and $ H_{10}^{u,d} $ acquire masses, as dictated by the model. They do so in such a way that all triplets are heavy, while only two light Higgs doublets remain at low scales, which we may associate with the MSSM Higgs doublets.
This is important because any light coloured Higgs states would lead to very rapid proton decay, so we need a justification for why certain doublet states remain light, while triplets remain heavy, a problem known as doublet-triplet splitting. We solve this \`a la Dimopolous-Wilczek \cite{DW}.
In $SO(10)$ there is a further complication, where each $ H_{10} $ has two $ SU(2) $ doublet states within it, and we have two additional doublets from the $ H_{16,\overbar{16}} $. Only two of these six doublets should be light, so as to reduce to the MSSM. We also show how we achieve this.

\subsubsection{Doublet-doublet splitting}
The superpotential that gives masses to doublets is
\begin{equation}
\begin{split}
	\mathcal{W}_{\mu} 
	&\sim 
	Z H_{10}^u H_{10}^u\frac{\xi^{6}}{M_Z^6}+Z H_{10}^u H_{10}^d\frac{\xi^{7}}{M_Z^7}+Z H_{10}^d H_{10}^d\frac{\xi^{8}}{M_Z^8} + \xi H_{16}H_{\overbar{16}}
	\\ &\quad 
	+ \frac{Z}{M_\Sigma}
	\left(H_{16}H_{16}H_{10}^d +\frac{\xi^8}{M_\Sigma^8}H_{16}H_{16}H_{10}^u
		+ H_{\overbar{16}}H_{\overbar{16}}H_{10}^u
		+ \frac{\xi}{M_\Sigma} H_{\overbar{16}}H_{\overbar{16}}H_{10}^d
	\right),
\label{eq:doubmass}
\end{split}
\end{equation}
where the corresponding diagrams that produce the nonrenormalisable terms are given in Fig.~\ref{fig:dtsplitting}. The second line is a reproduction of the second line in Eq.~\ref{eq:gutbreak}, while the first line includes those terms involving more than one insertion of $ H_{10}^{u,d} $, which had previously been omitted.

The fields $H_{16,\overbar{16}}$ contain doublets that mix with the ones in $H^{u,d}_{10}$, and will also contribute to masses, since they both get VEVs above the GUT scale. Defining 
$\tilde{H}_{16,\overbar{16}}= \braket{H_{16,\overbar{16}}}/M_\Sigma$,
we assume it to be reasonably close to 1.
We similarly define 
$\tilde{\xi} \sim \braket{\xi}/M_Z \sim \braket{\xi}/M_\Sigma $, where $ M_Z $ and $ M_\Sigma $ are the typical scales of the messengers that produce Eq.~\ref{eq:doubmass}.
We will find that they are necessarily different from the scale $ M_\zeta $ that governs the flavon driving potential. 

In order to make the connection to the two MSSM Higgs doublets, typically denoted $ H_u $ and $ H_d $ in reference to their behaviour under $ SU(2) $, 
we call the $H_u$-like doublets inside the $H^u_{10}, H^d_{10}$ and $ H_{\overbar{16}}$ fields, respectively, $H^u_u, H^d_u $ and $ H^{\overbar{16}}_u$. The $ H_d $-like doublets are named similarly, replacing the subindex $u \to d$. In other words,
\begin{equation}
\arraycolsep=1.4pt
\renewcommand{\arraystretch}{1.2}
	\begin{array}{rclrcl}
		\mathbf{2}_u(H^u_{10}) &\equiv& H_u^u, &\qquad \mathbf{2}_d(H_{10}^u) &\equiv& H_d^u, \\
		\mathbf{2}_u(H^d_{10}) &\equiv& H_u^d, & \mathbf{2}_d(H_{10}^d) &\equiv& H_d^d, \\
		\mathbf{2}_u(H_{\overbar{16}}) &\equiv& H_u^{\overbar{16}}, & \mathbf{2}_d(H_{\overbar{16}}) &\equiv& H_d^{\overbar{16}}.
	\end{array}
	\label{eq:doubdef}
\end{equation}

The doublet mass matrix can then be written in matrix form as
\begin{equation}
	M_D\sim \begin{blockarray}{c ccc}
	& H^u_u & H^d_u & H^{\overbar{16}}_u\\[1ex]
	\begin{block}{c(ccc)}
	H^u_d~ & \tilde{\xi}^{6} &  \tilde{\xi}^7& \tilde{H}_{\overbar{16}} \\[0.5ex]
	H^d_d & \tilde{\xi}^7 &  \tilde{\xi}^8 & \tilde{\xi}\tilde{H}_{\overbar{16}} \\[0.5ex]
	H^{16}_d & \tilde{H}_{16}\tilde{\xi}^8  & \tilde{H}_{16} & \xi/\braket{Z}\\
	\end{block}
	\end{blockarray}\braket{Z}
\label{eq:domass}
\end{equation}
Its eigenvalues $ m_D $ are
\begin{equation}
	m_D\sim \tilde{\xi}\braket{Z}, \quad \tilde{\xi}\braket{Z}, \quad \tilde{\xi}^8\braket{Z}.
\end{equation}
Two doublets receive large masses, which we assume are slightly larger than $M_{\rm GUT}$ such that they don't upset gauge coupling unification. The remaining eigenvalue is suppressed by a factor $ \tilde{\xi}^8 $. We can choose the mass of the $Z$ and $ \Sigma $ messengers so that $\tilde{\xi}\sim 0.03$, i.e. $ \tilde{\xi}^8 \braket{Z} \sim 1 $ TeV. This generates the MSSM $ \mu $ term $ \tilde{\xi}^8 \braket{Z} H_u^u H_d^d $ at the correct scale, where we make the connection to MSSM Higgs doublets by
\begin{equation}
	H_u \sim H^u_u,\qquad H_d\sim H^d_d.
\end{equation}

\subsubsection{Doublet-triplet splitting}
The Dimopoulos-Wilczek mechanism \cite{DW} is based on having an $SO(10)$ $\mathbf{45}$, which we call $H_{DW}$, that obtains a VEV with the structure
\begin{equation}
	\braket{H_{DW}}
	= \pmatr{0 & \braket{H_{U(5)}} \\ -\braket{H_{U(5)}} & 0},
\end{equation}
which is traceless regardless of the structure of $\braket{H_{U(5)}}$. We can actually choose $\braket{H_{U(5)}}=v_{45}~ \mathrm{diag}(1,1,1,0,0)$ such that it contributes only to the mass of the triplets \cite{DW}. This VEV alignment is not possible in an $SU(5)$ adjoint representation but is possible in the $SO(10)$ one.
The field $H_{DW}$ has an $R$-charge of 2 and a $\mathbb{Z}_9$ charge of 6, allowing us to write the term
\begin{equation}
	\mathcal{W}_{DT}\sim H_{DW}H_{10}^u H_{10}^d\frac{\xi}{M_\Upsilon},
	\label{eq:WDT}
\end{equation}
where, due to the antisymmetry of $ \braket{H_{DW}} $, only the mixed term is possible. The renormalisable diagram that produces this term is given in Fig. \ref{fig:dtsplitting}.

In analogy to Eq.~\ref{eq:doubdef}, we define Higgs triplets $ T $ arising from $ H^u_{10} $, $ H^d_{10} $ and $ H_{\overbar{16}} $ by
\begin{equation}
\arraycolsep=1.4pt
\renewcommand{\arraystretch}{1.2}
	\begin{array}{rclrcl}
		\mathbf{3}_u(H^u_{10}) &\equiv& T_u^u, &\qquad \mathbf{3}_d(H_{10}^u) &\equiv& T_d^u, \\
		\mathbf{3}_u(H^d_{10}) &\equiv& T_u^d, & \mathbf{3}_d(H_{10}^d) &\equiv& T_d^d, \\
		\mathbf{3}_u(H_{\overbar{16}}) &\equiv& T_u^{\overbar{16}}, & \mathbf{3}_d(H_{\overbar{16}}) &\equiv& T_d^{\overbar{16}}.
	\end{array}
\end{equation}
The terms involving these triplets arising from the superpotential in Eq.~\ref{eq:doubmass} produces the mass matrix
\begin{equation}
	M_T\sim \begin{blockarray}{c ccc}
	& T^u_u & T^d_u & T^{\overbar{16}}_u\\[1ex]
	\begin{block}{c(ccc)}
	T^u_d & \tilde{\xi}^{6} & \tilde{\xi}\braket{H_{DW}}/\braket{Z}& \tilde{H}_{\overbar{16}} \\[0.5ex]
	T^d_d &\tilde{\xi}\braket{H_{DW}}/\braket{Z} & \tilde{\xi}^8 & \tilde{\xi}\tilde{H}_{\overbar{16}} \\[0.5ex]
	T^{16}_d &\tilde{H}_{16}\,\tilde{\xi}^8  & \tilde{H}_{16} & \xi/\braket{Z}\\
	\end{block}
	\end{blockarray}\braket{Z},
\label{eq:trmass}
\end{equation}
where the only structural difference between this and Eq.~\ref{eq:domass} is in the (12) and (21) entries, which arise from Eq.~\ref{eq:WDT}. All the eigenvalues of this matrix are at the scale $\tilde{\xi}\braket{Z} \simlt M_{\rm GUT}$, i.e. there are no light triplet eigenstates, which gives doublet-triplet splitting.

\subsection{Proton decay}
A classic problem in GUT theories, including $ SO(10) $, is the prediction of excessively fast proton decay. The most dangerous processes come from the ``dimension 5'' operators $\Psi\Psi\Psi\Psi$ (for a discussion of dimension 6 operators we refer the reader to \cite{Antusch:2014poa}).
These ``dimension 5'' operators are forbidden by the symmetries of the model, but related higher-order operators of the following form are allowed:
\begin{equation}
	\Psi\Psi\Psi\Psi \frac{Z \overbar{\phi}}{M^3}\left(\frac{\xi}{M}\right)^{n},
\label{eq:protondecay}
\end{equation}
where $n$ is some positive integer.
Since we are working with the renormalisable theory, in order for this type of effective term to be present 
with $M\sim M_{\rm GUT}$, there would have to be GUT scale messengers allowing them. Specifically, to produce the above term we would need messengers that are $ \Delta(27) $ triplets, which are completely absent from our model. Hence such terms can not be produced at (or below) the GUT scale.

Such operators may, however, arise with Planck scale suppression, i.e. $ M \sim M_P $. 
Specifically the lowest order term arising from fields that aquires non-vanishing VEVs and therefore contributes to proton decay in our model is
\begin{equation}
	\Psi\Psi\Psi\Psi \frac{Z \overbar{\phi}_{\rm dec} \xi^3}{M^6_P},
\end{equation}
which would generate dangerous proton decay terms of the type
\begin{equation}
gQQQL\frac{\braket{X}}{M_P^2},
\end{equation}
where $g$ is a dimensionless coupling and $\braket{X}$ is a generic VEV of a field, as discussed in \cite{Murayama:1994tc}. These terms must be suppressed enough to generate a proton lifetime $\tau_p >10^{32} \mathrm{~yrs}$, which is achieved when
\begin{equation}
	g\braket{X}<3\times 10^{9} \mathrm{~GeV}.
\end{equation}

In our model,
\begin{equation}
\braket{X}=\frac{\braket{Z} v_{\rm dec} \braket{\xi}^3}{M^4_P}\sim 150 \mathrm{~GeV},
\end{equation}
such that with $\mathcal{O}(1)$ dimensionless couplings, it predicts very suppressed proton decay.

\section{Numerical fit}
\label{sec:fit}

\subsection{Low-scale mass matrices}
At the low scale, the VEVs of flavons and messenger fields combine to give the following 
mass matrices:
\addtocounter{equation}{1}
\begingroup
\newcommand{\pmatrt}[1]{\hskip -1pt\begin{pmatrix} #1 \end{pmatrix}}
\newcommand{\pmatrs}[1]{\hskip -2pt\mbox{\smaller$\begin{pmatrix} #1 \end{pmatrix}$}}
\begin{empheq}{alignat=8}
	m^u &= v^u \Bigg[ y^u_1 &\pmatrs{0 & 0 & 0 \\ 0 & 1 & 1 \\ 0 & 1 & 1}
		&+ y^u_2 e^{i \eta^u_2} &\pmatrs{1 & 3 & 1 \\ 3 & 9 & 3 \\ 1 & 3 & 1}
		&+ y^u_3 e^{i \eta^u_3} &\pmatrs{0 & 0 & 0 \\ 0 & 0 & 0 \\ 0 & 0 & 1} 
		&+ y^u_4 e^{i \eta^u_4} &\pmatrs{0 & 0 & 1 \\ 0 & 0 & 3 \\ 1 & 3 & 2}\Bigg]
		\tag{\theequation a}\\
	m^d &= v^d \Bigg[ y^d_1 &\pmatrs{0 & 0 & 0 \\ 0 & 1 & 1 \\ 0 & 1 & 1}
		&+ y^d_2 e^{i \eta^d_2} &\pmatrs{1 & 3 & 1 \\ 3 & 9 & 3 \\ 1 & 3 & 1}
		&+ y^d_3 e^{i \eta^d_3} &\pmatrs{0 & 0 & 0 \\ 0 & 0 & 0 \\ 0 & 0 & 1} 
		&\Bigg]&
		\tag{\theequation b}\\
	m^e &= v^d \Bigg[y^e_1 &\pmatrs{0 & 0 & 0 \\ 0 & 1 & 1 \\ 0 & 1 & 1}
		&+ y^e_2 e^{i \eta} &\pmatrs{1 & 3 & 1 \\ 3 & 9 & 3 \\ 1 & 3 & 1}
		&+ y^e_3 e^{i \eta^\prime} &\pmatrs{0 & 0 & 0 \\ 0 & 0 & 0 \\ 0 & 0 & 1}
		&\Bigg]&
		\tag{\theequation c}\\
	m^\nu &= \hspace{3ex}\mu_a &\pmatrs{0 & 0 & 0 \\ 0 & 1 & 1 \\ 0 & 1 & 1}
		&+ \mu_b e^{i\eta} &\pmatrs{1 & 3 & 1 \\ 3 & 9 & 3 \\ 1 & 3 & 1}
		&+ \mu_c e^{i\eta^\prime} &\pmatrs{0 & 0 & 0 \\ 0 & 0 & 0 \\ 0 & 0 & 1}
		& & \tag{\theequation d}
\end{empheq}
\endgroup
where $ v^u = v \sin \beta $ and $ v^d = v \cos \beta $ are the VEVs of the MSSM Higgs fields, and $ v = 174 $ GeV.
We recall from Eq.~\ref{eq:physicalphases} that $ \eta = 2\pi/3 $, while $ \eta^\prime = 0 $, while the remaining phases are free. 

Assuming all superpotential terms have $ \mathcal{O}(1) $ couplings, we may derive a ``natural'' scale for each of the coefficients $ y^f_i $. Firstly, we recall that there are several messenger scales present in our model. The ones that appear in $ y^f_i $ are $ M_\chi $, $ M_\zeta $, $ M_{\Omega_{\rm dec}} $, $ M_{\Omega_{\rm sol}} $ and $ M_{\Omega_{\rm atm}} $. 
As previously established, we have $ \braket{\xi} \simlt M_\zeta < M_\chi $. More specifically, we will assume the following ratios:
\begin{equation}
	\frac{\braket{\xi}}{M_\zeta} \simgt 0.5, \qquad
	\frac{\braket{\xi}}{M_\chi} \simlt 0.1.
\label{eq:xibounds}
\end{equation}
We further define the GUT scale by $ M_{\rm GUT} \equiv v_{45} \simlt M_\chi $.
Finally, as discussed previously, we assume that $ M_{\Omega_{\rm atm}} \approx M_{\Omega_{\rm sol}} > M_{\Omega_{\rm dec}}$, by roughly one order of magnitude.

The coefficients $ y^f_i $ derive from terms in in Eq.~\ref{eq:sYW}, which take a generic form
\begin{equation}
\begin{split}
y^f_1 &= \overbar{\phi}_{\rm atm} \overbar{\phi}_{\rm atm} \xi^{N-2} \sum_{n=0}^N \frac{\lambda^{(f)}_{X,n}}{\braket{H_{45}}^n M_{\chi}^{N-n}},\\
y^f_2 &= \overbar{\phi}_{\rm sol} \overbar{\phi}_{\rm sol} \xi^{N-2} \sum_{n=0}^N \frac{\lambda^{(f)}_{X,n}}{\braket{H_{45}}^n M_{\chi}^{N-n}},\\
y^f_3 &= \overbar{\phi}_{\rm dec} \overbar{\phi}_{\rm dec} \xi^{N-2} \sum_{n=0}^N \frac{\lambda^{(f)}_{X,n}}{\braket{H_{45}}^n M_{\chi}^{N-n}},
\end{split}
\label{eq:yfi}
\end{equation}
where $ \lambda $ are $ \mathcal{O}(1) $ couplings and $ N $ is a number between two and five. We will assume there are no large cancellations between terms in the sums.
The flavon VEVs are discussed in Section \ref{sec:driving}, from which we may approximate their VEVs by
\begin{align}
	\braket{\overbar{\phi}_{\rm dec}} \sim M_{\rm GUT} ,\quad
	\braket{\overbar{\phi}_{\rm atm}} \sim \frac{\braket{\xi}^4}{M_\zeta^4} M_{\rm GUT}
	,\quad
	\braket{\overbar{\phi}_{\rm sol}} \sim \frac{\braket{\xi}^{7/2}}{M_\zeta^{7/2}} M_{\rm GUT}.
\end{align}
We note immediately that these VEVs have large powers of $ \braket{\xi}/M_{\zeta} $, which is primarily bounded below (see Eq.~\ref{eq:xibounds}). This translates to only a loose upper bound on the fitting parameters.
From Eqs.~\ref{eq:xibounds}-\ref{eq:yfi}, we expect the following scales for the fitted coefficients:
\begin{alignat}{4}
	y^u_1 &\simgt 4\times10^{-4}, \qquad&
	y^e_1 &\simgt 4\times10^{-5}, \qquad&
	y^d_1 &\simgt 4\times10^{-5}, \qquad& 
	\mu_a &\sim 10^{-2} \mathrm{~eV}, \nonumber\\
	y^u_2 &\simgt 8\times10^{-5}, &
	y^e_2 &\simgt 8\times10^{-6}, &
	y^d_2 &\simgt 8\times10^{-6}, &
	\mu_b &\sim 10^{-3} \mathrm{~eV},\nonumber\\
	y^u_3 &\sim 1, &
	y^e_3 &\sim 10^{-1}, &
	y^d_3 &\sim 10^{-1}, &
	\mu_c &\sim 10^{-3} \mathrm{~eV}, \nonumber\\
	y^u_4 &\simgt 5\times10^{-4}. & &
\end{alignat}

\subsection{Fitting procedure}\label{sec:fitproc}
To fit the real coefficients $y^u_{1,2,3,4}$, $y^d_{1,2,3}$, $y^e_{1,2,3}$, and $\mu_{a,b,c}$ as well as phases $ \eta^u_{2,3,4} $ and $ \eta^d_{2,3} $,
we wish to minimise a $\chi^2$ function that relates the $N$ physical predictions $P_i(\{x\})$ for a given set of input parameters $\{x\}$ to their current best-fit values $\mu_i$ and their associated $1\sigma$ errors, denoted $ \sigma_i $. It is defined by
\begin{equation}
	\chi^2 = \sum_{i=1}^N \left(\frac{P_i(\{x\})-\mu_i}{\sigma_i} \right)^2.
	\label{eq:chisq}
\end{equation}
The errors $ \sigma_i $ are equivalent to the standard deviation of the experimental fits to a Gaussian distribution. For most parameters, their distribution is essentially Gaussian, with the exception of the (lepton) atmospheric angle $ \theta^l_{23} $. 

For a normal hierarchy (as predicted by the model), the distribution is roughly centered on maximal atmospheric angle, i.e. $ (\theta^l_{23})^{\rm best-fit} \sim 45^\circ $, while the best fit value is given by $ \theta^l_{23} = 42.3^\circ $, i.e. there is a small preference for $ \theta_{23}^l $ to be in the first octant. 
As such, there are two possible scenarios to consider when performing our fit.
\vspace{-3ex}
\begin{itemize}
\renewcommand{\labelitemi}{$\circ$}
\renewcommand{\itemsep}{-1ex}
	\item Scenario 1: we assume that the (weak) preference for $ \theta^l_{23} < 45^\circ $ is true, and approximate its distribution by a Gaussian about $ 42.3^\circ $, setting $ \sigma_{\theta^l_{23}} = 1.6^\circ $ as the error.
	\item Scenario 2: we remain octant-agnostic by assuming a Gaussian distribution centred at the midpoint between the two $ 1\sigma $ bounds, i.e. $(\theta^l_{23})^{\rm best-fit} = 45.9^\circ $ with $ \sigma_{\theta^l_{23}} = 3.5^\circ $.
\end{itemize}
\vspace{-4ex}
Below we present the results of our fit in each of these two scenarios.

In this analysis, $N = 18$, corresponding to six mixing angles $\theta^l_{ij}$ (neutrinos) and $\theta^q_{ij}$ (quarks), the CKM phase $\delta^q$, nine Yukawa eigenvalues for the quarks and charged leptons, and two neutrino mass-squared differences $\Delta m^2_{21}$ and $\Delta m^2_{31}$. 
In the lepton sector, we use the PDG parametrisation of the PMNS matrix \cite{Beringer:1900zz}
\(U_{\mathrm{PMNS}} = R^l_{23} U^l_{13} R^l_{12} P_\textrm{PDG} \) in terms of 
\(s_{ij}=\sin \theta^l_{ij}\), \(c_{ij}=\cos\theta^l_{ij}\), the Dirac CP violating phase \(\delta^l\) and further Majorana phases contained in 
\(P_\textrm{PDG} = \textrm{diag}(1,e^{i\frac{\alpha_{21}}{2}},e^{i\frac{\alpha_{31}}{2}})\). 
Experimentally, the leptonic phase $ \delta^l $ is poorly constrained at $ 1\sigma $ (and completely unconstrained at $ 3\sigma $), so is not fit, and left as a pure prediction of the model, as are the (completely unconstrained) Majorana phases $ \alpha_{21} $ and $ \alpha_{31} $. 

The running of best-fit and error values to the GUT scale are generally dependent on SUSY parameters, notably $\tan \beta$, as well as contributions from SUSY threshold corrections. 
We extract the GUT scale CKM parameters and all Yukawa couplings (with associated errors) from \cite{Antusch:2013jca} for $ \tan \beta = 5 $. The value of $ \tan \beta $ does not have a significant impact on the quality of our model, so we only present results for $ \tan \beta = 5 $ here.
We find that our model is essentially unaffected by threshold corrections, so we simply assume them to be zero. In further reference to \cite{Antusch:2013jca}, this is equivalent to setting the parameters $ \bar{\eta}_i $ to zero.
Experimental neutrino parameters are extracted from \cite{Gonzalez-Garcia:2014bfa}. 

\begin{table}[!ht]
\centering
\footnotesize
\renewcommand{\arraystretch}{1.3}
\begin{minipage}[b]{0.50\textwidth}
\centering
\captionsetup{width=0.9\textwidth}
\begin{tabular}{| @{\hskip 8pt}c@{\hskip 6pt}l | c | r@{\hskip 3pt}c@{\hskip 3pt}l |}
\hline
\multicolumn{2}{|c|}{Observables} & \rule{0pt}{4.5ex}Model & \multicolumn{3}{c|}{\makecell{Data fit $ 1\sigma $ range \\ (from \cite{Antusch:2013jca})}} \\[2ex]
\hline\hline
\rule{0pt}{3ex}%
	$ \theta_{12}^q $ & /\degr & 13.024 & 12.985 &$\rightarrow$& 13.067 \\
	$ \theta_{13}^q $ & /\degr & 0.1984 & 0.1866 &$\rightarrow$& 0.2005 \\
	$ \theta_{23}^q $ & /\degr & 2.238 & 2.202 &$\rightarrow$& 2.273 \\
	$ \delta^q $ & /\degr & 69.32 & 66.12 &$\rightarrow$& 72.31 \\
\rule{0pt}{3ex}%
	$ m_{u} $ & /MeV & 0.575 & 0.351 &$\rightarrow$& 0.666 \\
	$ m_{c} $ & /MeV & 248.4 & 240.1 &$\rightarrow$& 257.5 \\
	$ m_{t} $ & /GeV & 92.79 & 89.84 &$\rightarrow$& 95.77 \\
	$ m_{d} $ & /MeV & 0.824 & 0.744 &$\rightarrow$& 0.929 \\
	$ m_{s} $ & /MeV & 15.55 & 15.66 &$\rightarrow$& 17.47 \\
	$ m_{b} $ & /GeV & 0.939 & 0.925 &$\rightarrow$& 0.948 \\[0.5ex]
\hline
	\multicolumn{2}{|c|}{\rule{0pt}{3ex}$ \delta\chi^2 $} & 2.0 &&&\multicolumn{1}{c}{} \\[0.5ex]
\cline{1-3}
\end{tabular}
\caption{Model predictions in the quark sector, for $ \tan \beta = 5 $. The quark contribution to the total $ \chi^2 $ is 2.0. The observables are at the GUT scale.}
\label{tab:fitoutq}
\end{minipage}%
\qquad%
\begin{minipage}[b]{0.4\textwidth}
\captionsetup{width=0.8\textwidth}
\centering
\begin{tabular}{| c | c |}
\hline
\rule{0pt}{4.5ex}%
	Parameter	& Fitted value \\[2ex]
\hline\hline
\rule{0pt}{3ex}%
	$ y^u_1 $ & 3.314 \exto{-5} \\
	$ y^u_2 $ & 2.060 \exto{-4} \\
	$ y^u_3 $ & 5.503 \exto{-1} \\
	$ y^u_4 $ & 7.423 \exto{-3} \\
\rule{0pt}{3ex}%
	$ \eta^u_2 $ & $ 0.617\pi $ \\
	$ \eta^u_3 $ & $ 1.047\pi $ \\
	$ \eta^u_4 $ & $ 1.718\pi $ \\
\rule{0pt}{3ex}%
	$ y^d_1 $ & 3.288 \exto{-4} \\
	$ y^d_2 $ & 3.308 \exto{-5} \\
	$ y^d_3 $ & 2.785 \exto{-2} \\
\rule{0pt}{3ex}%
	$ \eta^d_2 $ & $ 0.521\pi $ \\
	$ \eta^d_3 $ & $ 1.065\pi $ \\[0.5ex]
\hline
\end{tabular}
\caption{Quark sector input parameter values.\\ \\}
\label{tab:fitinq}
\end{minipage}
\end{table}

\begin{table}[!ht]
\centering
\footnotesize
\renewcommand{\arraystretch}{1.3}
\begin{tabular}{| c@{\hskip 4pt}l | cc | r@{\hskip 3pt}c@{\hskip 3pt}l |}
\hline
\multicolumn{2}{|c|}{\multirow{2}{*}{\rule{0pt}{6ex}Observables}} & \multicolumn{2}{c|}{\rule{0pt}{3.5ex}Model} & \multicolumn{3}{c|}{\multirow{2}{*}{\rule{0pt}{6.5ex}\makecell{Data fit $ 1\sigma $ range \\ (from \cite{Antusch:2013jca,Gonzalez-Garcia:2014bfa,Ade:2015xua})}}} \\[1ex]
\cline{3-4}
&& \rule{0pt}{4.5ex}\makecell{Scenario 1\\($ \theta^{\rm exp,bf}_{23} = 42.3^\circ $)} & \makecell{Scenario 2\\($ \theta^{\rm exp,bf}_{23} = 45.9^\circ $)} &&&\\[2ex]
\hline\hline
\rule{0pt}{3ex}%
	$ \theta_{12}^l $ & /\degr & 33.13 & 32.94 & 32.83 &$\rightarrow$& 34.27 \\
	$ \theta_{13}^l $ & /\degr & 8.59 & 8.55 & 8.29 &$\rightarrow$& 8.68 \\
\rule{0pt}{3ex}%
	\multirow{2}{*}{$ \theta_{23}^l $} & \multirow{2}{*}{/\degr} & 40.81 &  & 40.63 &$\rightarrow$& 43.85 \\
	 & &  & 46.65 & 42.40 &$\rightarrow$& 49.40 \\
\rule{0pt}{3ex}%
	$ \delta^l $ & /\degr & 280 & 275 & 192 &$\rightarrow$& 318 \\
\rule{0pt}{3ex}%
	$ m_{e} $ & /MeV & 0.342 & 0.342 & 0.340 & $\rightarrow$& 0.344 \\
	$ m_{\mu} $ & /MeV & 72.25 & 72.25 & 71.81 & $\rightarrow$& 72.68 \\
	$ m_{\tau} $ & /GeV & 1.229 & 1.229 & 1.223 & $\rightarrow$& 1.236 \\
\rule{0pt}{3ex}%
	$ \Delta m_{21}^2 $ & /eV$^2$ & 7.58 \exto{-5} & 7.46 \exto{-5} & (7.33 & $\rightarrow$& 7.69) \exto{-5} \\
	$ \Delta m_{31}^2 $ & /eV$^2$ & 2.44 \exto{-3} & 2.47 \exto{-3} & (2.41 & $\rightarrow$& 2.50) \exto{-3} \\
\rule{0pt}{3ex}%
	$ m_1 $ & /meV & 0.32 & 0.38 & & $ - $ & \\
	$ m_2 $ & /meV & 8.64 & 8.65 & & $ - $ & \\
	$ m_3 $ & /meV & 49.7 & 49.7 & & $ - $ & \\
	$ \sum m_i $ & /meV & 58.7 & 59.4 &  &$ < $& 230 \\
\rule{0pt}{3ex}%
	$ \alpha_{21} $ & /\degr & 264 & 264 & & $ - $ & \\
	$ \alpha_{31} $ & /\degr & 323 & 333 & & $ - $ & \\
	$ | m_{ee} | $ & /meV & 2.46 & 2.42 & & $-$ & \\[0.5ex]
\hline
	\multicolumn{2}{|c|}{\rule{0pt}{3ex}$ \delta\chi^2 $} & 1.3 & 0.7 &&&\multicolumn{1}{c}{} \\[0.5ex]
\cline{1-4}
\end{tabular}
\caption{Model predictions in the lepton sector, for $ \tan \beta = 5 $. The observables are at the GUT scale. The lepton contributions to the total $ \chi^2 $ are 1.3 and 0.7 in scenario 1 and 2, respectively. Note the two different data fit $ 1\sigma $ ranges for $ \theta_{23}^l $, depending on the choice of scenario, as discussed in Section \ref{sec:fitproc}.}
\label{tab:fitoutl}
\end{table}

\begin{table}
\centering
\footnotesize
\renewcommand{\arraystretch}{1.3}
\begin{tabular}{| c | cc |}
\hline
	\multirow{2}{*}{\rule{0pt}{5ex}Parameter}	& \multicolumn{2}{c|}{\rule{0pt}{3.5ex}Fitted value} \\[1ex]
\cline{2-3}
& \rule{0pt}{4.5ex}\makecell{Scenario 1\\($ \theta^{\rm exp,bf}_{23} = 42.3^\circ $)} & \makecell{Scenario 2\\($ \theta^{\rm exp,bf}_{23} = 45.9^\circ $)} \\[2ex]
\hline\hline
\rule{0pt}{3ex}%
	$ y^e_1 $ & 2.217 \exto{-3} & -1.966 \exto{-3} \\
	$ y^e_2 $ & -1.025 \exto{-5} & 1.027 \exto{-5} \\
	$ y^e_3 $ & 3.366 \exto{-2} & 3.790 \exto{-2} \\
\rule{0pt}{3ex}%
	$ \mu_a $ /meV & 26.60 & 25.90 \\
	$ \mu_b $ /meV & 2.571 & 2.546 \\
	$ \mu_c $ /meV & 2.052 & 2.461 \\
\rule{0pt}{3ex}%
	$ \eta $ & \multicolumn{2}{c|}{$ 2\pi/3 $} \\
	$ \eta^\prime $ & \multicolumn{2}{c|}{0} \\ [0.5ex]
\hline
\end{tabular}
\caption{Lepton input parameter values (with $\eta,\eta'$ fixed by the theory).}
\label{tab:fitinl}
\end{table}

Tables \ref{tab:fitoutq}-\ref{tab:fitinq} show the numerical fit of all relevant parameters to quark mass and mixing data, while Tables \ref{tab:fitoutl}-\ref{tab:fitinl} show the fit to lepton mass and mixing. The fit gives $ \chi^2 \approx 3.3 $ and $ \chi^2 \approx 2.7 $ in scenarios 1 and 2, respectively.

Although the model is not technically predictive in the quark sector due to an excess of free parameters, the structure of the mass matrices forces some tension between parameters, notably the $ \sim 1 \sigma $ deviation from the experimental fit value in the strange quark mass.

In the lepton sector there are two fixed discrete phases plus six continuous input parameters that we fit to three charged lepton masses, two neutrino mass-squared differences and three mixing angles (a total of eight observables), while predicting the CP phase $ \delta^l $, two Majorana phases and the effective neutrino mass $ |m_{ee}| $.
Although our fit does not constitute a full analysis of the parameter space, it agrees with the results of a more dedicated numerical analysis of CSD($n$) models \cite{Bjorkeroth:2014vha}.

\FloatBarrier
\section{Conclusion}
\label{sec:conclusion}

We have proposed a renormalisable $\Delta(27)\times SO(10)$  SUSY GUT of Flavour.
All symmetries, including an additional $\mathbb{Z}_{9} \times \mathbb{Z}_{12} \times \mathbb{Z}_{4}^{R}$ discrete symmetry, are broken close to the GUT breaking scale due to the action of explicit superpotential terms to yield
the MSSM with the standard R-parity as the surviving theory at low energies.

The model is very ambitious since it is not only a full $SO(10)$ SUSY GUT theory, with GUT symmetry breaking sectors including doublet-triplet splitting and guaranteeing the absence of extra light doublets, but it also addresses the flavour problem due to additional commuting discrete family symmetries.
The mystery of why there are three families of quarks and leptons including their observed pattern of masses and mixing angles is addressed, and a novel form of spontaneous geometrical CP violation arises from the nature of the $\Delta(27)$ group. 

In many respects $SO(10)$ is the ``holy grail'' of GUT groups since it involves probably the most elegant unification of quarks and leptons, including a right-handed neutrino, making neutrino mass and mixing inevitable (unlike $SU(5)$ where neutrino masses could quite happily be set to zero). When combined with the family symmetry $\Delta(27)$, all quarks and leptons are unified into a single multiplet $(3,16)$ providing a very elegant and simple unification of all matter. Such a complete flavour unification has been attempted before,
but until now the technicalities involved have led to only partial success. The contribution of the present paper lies in showing how many of these technical difficulties may be successfully overcome within a fully fledged 
$\Delta(27)\times SO(10)$  SUSY GUT of Flavour.

We emphasise that in our model all quark and lepton (including neutrino) mass matrices take a particularly simple universal form,
with a small correction to the up-type quark mass matrix being responsible for quark mixing. 
The heavy right-handed neutrino Majorana matrix also has the same universal form, and even including the see-saw mechanism, the low energy effective light left-handed Majorana neutrino mass matrix also has this form,
indeed corresponding to CSD3, leading to a highly predictive scheme for leptonic mixing.
The model predicts a normal neutrino mass hierarchy with the best-fit lightest neutrino mass $ m_1 \approx 0.32 $ or $ m_1 \approx 0.38$ meV,
and all neutrino parameters fitted to within $ 1\sigma $ of the values predicted by global fits to experiment. In particular, we predict a CP-violating oscillation phase $ \delta^l \approx 280^\circ $ or $ \delta^l \approx 275^\circ$, in agreement with current experimental hints.

The model has the following virtues which we summarise as follows:
\vspace{-2ex}
\begin{itemize}
\renewcommand{\labelitemi}{$\circ$}
\renewcommand{\itemsep}{-1ex}
\item It is fully renormalisable at the GUT scale, with an explicit $SO(10)$ breaking sector and a spontaneously broken CP symmetry.  
\item It involves only the smaller ``named'' representations of $SO(10)$.
\item The MSSM is reproduced below the GUT scale, with R-parity emerging from a discrete $\mathbb{Z}_4^R$.
\item Doublet-triplet splitting is achieved through the Dimopoulos-Wilczek mechanism.
\item A $ \mu $ term is generated at the correct scale.
\item Proton decay is sufficiently suppressed.
\item $ \Delta(27) $ justifies the CSD3 alignment.
\item Spontaneous geometrical CP violation, where the input phase is the cube root of unity $\omega = e^{2i \pi/3}$, originates from the $ \Delta(27) $.
\item We successfully fit all quark and lepton masses, with the PMNS mixing matrix predicted 
(with no free parameters) once the physical neutrino masses are specified.
\end{itemize}
These features are desirable for any flavour or GUT model. Achieving them all in the same model represents a significant step towards a complete flavoured $SO(10)$ GUT.
At the cost of its large field content, the model is rather successful and fairly complete. 
Nevertheless, some relevant topics remain beyond the scope of the paper.
Notably, while we discuss how the model leads to an effective MSSM after the symmetries are broken, we do not discuss the details of SUSY breaking nor the mass spectrum of SUSY
partners at low energy.
Furthermore, we have not considered GUT threshold corrections and their effect at high energy on gauge coupling unification, nor a possible string theory completion for this model.

\section*{Acknowledgements}
This project has received funding from the European Union's Seventh Framework Programme for research, technological development and demonstration under grant agreement no PIEF-GA-2012-327195 SIFT.
The authors also acknowledge partial support from the European Union FP7 ITN-INVISIBLES (Marie Curie Actions, PITN- GA-2011- 289442) and CONACyT.


\begin{thebibliography}{99}

\bibitem{King:2013eh}
  S.~F.~King and C.~Luhn,
  Rept.\ Prog.\ Phys.\  {\bf 76} (2013) 056201
  [arXiv:1301.1340];
  S.~F.~King, A.~Merle, S.~Morisi, Y.~Shimizu and M.~Tanimoto,
  New J.\ Phys.\  {\bf 16} (2014) 045018
  [arXiv:1402.4271];
  S.~F.~King,
  J.\ Phys.\ G {\bf 42} (2015) 12,  123001
  [arXiv:1510.02091 [hep-ph]].

\bibitem{Antusch:2009gu}
  S.~Antusch and M.~Spinrath,
  Phys.\ Rev.\ D {\bf 79} (2009) 095004
  [arXiv:0902.4644 [hep-ph]];
  S.~Antusch, S.~F.~King and M.~Spinrath,
  Phys.\ Rev.\ D {\bf 89} (2014) 5,  055027
  [arXiv:1311.0877 [hep-ph]].
  
\bibitem{Antusch:2014poa}
  S.~Antusch, I.~de Medeiros Varzielas, V.~Maurer, C.~Sluka and M.~Spinrath,
  JHEP {\bf 1409} (2014) 141
  [arXiv:1405.6962 [hep-ph]].
  
\bibitem{Bjorkeroth:2015ora}
  F.~Bj\"orkeroth, F.~J.~de Anda, I.~de Medeiros Varzielas and S.~F.~King,
  JHEP {\bf 1506} (2015) 141
  [arXiv:1503.03306 [hep-ph]].

\bibitem{Fritzsch:1974nn}
  H.~Fritzsch and P.~Minkowski,
  Annals Phys.\  {\bf 93} (1975) 193.

\bibitem{Branco:1983tn} 
  G.~C.~Branco, J.~M.~Gerard and W.~Grimus,
  Phys.\ Lett.\ B {\bf 136}, 383 (1984);
  I.~de Medeiros Varzielas and D.~Emmanuel-Costa,
  Phys.\ Rev.\ D {\bf 84}, 117901 (2011)
  [arXiv:1106.5477 [hep-ph]];
  I.~de Medeiros Varzielas, D.~Emmanuel-Costa and P.~Leser,
  Phys.\ Lett.\ B {\bf 716}, 193 (2012)
  [arXiv:1204.3633 [hep-ph]];
  I.~de Medeiros Varzielas,
  JHEP {\bf 1208}, 055 (2012)
  [arXiv:1205.3780 [hep-ph]];
  G.~Bhattacharyya, I.~de Medeiros Varzielas and P.~Leser,
  Phys.\ Rev.\ Lett.\  {\bf 109}, 241603 (2012)
  [arXiv:1210.0545 [hep-ph]];
  I.~P.~Ivanov and L.~Lavoura,
  Eur.\ Phys.\ J.\ C {\bf 73}, no. 4, 2416 (2013)
  [arXiv:1302.3656 [hep-ph]];
 I.~de Medeiros Varzielas and D.~Pidt,
  J.\ Phys.\ G {\bf 41}, 025004 (2014)
  [arXiv:1307.0711 [hep-ph]];
  I.~Medeiros Varzielas and D.~Pidt,
  JHEP {\bf 1311}, 206 (2013)
  [arXiv:1307.6545 [hep-ph], arXiv:1307.6545];
  M.~Fallbacher and A.~Trautner,
  arXiv:1502.01829 [hep-ph].


\bibitem{DW}
S. Dimopoulos and F. Wilczek, Report No. NSF-ITP-82-07 (unpublished);
 K.~S.~Babu and S.~M.~Barr,
  Phys.\ Rev.\ D {\bf 48} (1993) 5354
  [hep-ph/9306242];
   S.~M.~Barr and S.~Raby,
  Phys.\ Rev.\ Lett.\  {\bf 79} (1997) 4748
  [hep-ph/9705366].

   
\bibitem{Minkowski:1977sc}
  P.~Minkowski,
  Phys.\ Lett.\  B {\bf 67} (1977) 421;
T. Yanagida, in Proceedings of the Workshop on Unified Theory and Baryon Number
of the Universe, eds. O. Sawada and A. Sugamoto (KEK, 1979) p.95;
 M. Gell-Mann,
P. Ramond and R. Slansky, in Supergravity, eds. P. van Niewwenhuizen and D.
Freedman (North Holland, Amsterdam, 1979) Conf.Proc. C790927 p.315, PRINT-80-0576.

\bibitem{Ramond:1979py}
  P.~Ramond, 
Invited talk given at Conference: C79-02-25
(Feb 1979) p.265-280, CALT-68-709,
  hep-ph/9809459.
           
  
\bibitem{Lee:2011dya}
  H.~M.~Lee, S.~Raby, M.~Ratz, G.~G.~Ross, R.~Schieren, K.~Schmidt-Hoberg and P.~K.~S.~Vaudrevange,
  Nucl.\ Phys.\ B {\bf 850} (2011) 1
  [arXiv:1102.3595 [hep-ph]].

\bibitem{King:1998jw}
  S.~F.~King,
  Phys.\ Lett.\  B {\bf 439} (1998) 350
  [hep-ph/9806440];\,
  S.~F.~King,
  Nucl.\ Phys.\  B {\bf 562} (1999) 57
  [hep-ph/9904210];\,
  S.~F.~King,
  Nucl.\ Phys.\  B {\bf 576} (2000) 85
  [hep-ph/9912492];\,
  S.~F.~King,
  JHEP {\bf 0209} (2002) 011
  [hep-ph/0204360];
  S.~F.~King,
  JHEP {\bf 0508} (2005) 105
  [hep-ph/0506297];
   S.~Antusch, S.~F.~King, C.~Luhn and M.~Spinrath,
  Nucl.\ Phys.\ B {\bf 856} (2012) 328
  [arXiv:1108.4278 [hep-ph]].
   
\bibitem{King:2013iva}
  S.~F.~King,
  JHEP {\bf 1307} (2013) 137
  [arXiv:1304.6264 [hep-ph]];
  %
\bibitem{Bjorkeroth:2014vha}
  F.~Bj\"orkeroth and S.~F.~King,
  J.\ Phys.\ G {\bf 42} (2015) 12,  125002
  [arXiv:1412.6996 [hep-ph]].
  
\bibitem{King:2013hoa}
  S.~F.~King,
  JHEP {\bf 1401} (2014) 119
  [arXiv:1311.3295 [hep-ph]].

  \bibitem{huge}
  S.~F.~King and G.~G.~Ross,
  Phys.\ Lett.\ B {\bf 520} (2001) 243
  [hep-ph/0108112];
  %
    S.~F.~King and G.~G.~Ross,
  Phys.\ Lett.\ B {\bf 574} (2003) 239
  [hep-ph/0307190];
  %
  I.~de Medeiros Varzielas and G.~G.~Ross,
  Nucl.\ Phys.\ B {\bf 733} (2006) 31
  [hep-ph/0507176];
  %
    I.~de Medeiros Varzielas, S.~F.~King and G.~G.~Ross,
  Phys.\ Lett.\ B {\bf 644} (2007) 153
  [hep-ph/0512313];
  S.~F.~King and M.~Malinsky,
  JHEP {\bf 0611} (2006) 071
  [hep-ph/0608021];
  G.~Altarelli, F.~Feruglio and C.~Hagedorn,
  JHEP {\bf 0803} (2008) 052
  [arXiv:0802.0090 [hep-ph]];
  P.~Ciafaloni, M.~Picariello, E.~Torrente-Lujan and A.~Urbano,
  Phys.\ Rev.\ D {\bf 79} (2009) 116010
  [arXiv:0901.2236 [hep-ph]];
   T.~J.~Burrows and S.~F.~King,
  Nucl.\ Phys.\ B {\bf 835} (2010) 174
  [arXiv:0909.1433 [hep-ph]];
  I.~K.~Cooper, S.~F.~King and C.~Luhn,
  Phys.\ Lett.\ B {\bf 690} (2010) 396
  [arXiv:1004.3243 [hep-ph]];
   S.~Antusch, S.~F.~King and M.~Spinrath,
  Phys.\ Rev.\ D {\bf 83} (2011) 013005
  [arXiv:1005.0708 [hep-ph]];
  S.~Antusch, S.~F.~King, C.~Luhn and M.~Spinrath,
  Nucl.\ Phys.\ B {\bf 850} (2011) 477
  [arXiv:1103.5930 [hep-ph]];
  D.~Meloni,
  JHEP {\bf 1110} (2011) 010
  [arXiv:1107.0221];
  T.~J.~Burrows and S.~F.~King,
  Nucl.\ Phys.\ B {\bf 842} (2011) 107
  [arXiv:1007.2310 [hep-ph]];
    I.~de Medeiros Varzielas,
  JHEP {\bf 1201} (2012) 097
  [arXiv:1111.3952 [hep-ph]];
  C.~Hagedorn, S.~F.~King and C.~Luhn,
  Phys.\ Lett.\ B {\bf 717} (2012) 207
  [arXiv:1205.3114];
    B.~D.~Callen and R.~R.~Volkas,
  Phys.\ Rev.\ D {\bf 86} (2012) 056007
  [arXiv:1205.3617];
  A.~Meroni, S.~T.~Petcov and M.~Spinrath,
  Phys.\ Rev.\ D {\bf 86} (2012) 113003
  [arXiv:1205.5241 [hep-ph]];
  S.~F.~King, C.~Luhn and A.~J.~Stuart,
  Nucl.\ Phys.\ B {\bf 867} (2013) 203
  [arXiv:1207.5741];
   S.~F.~King,
    Phys.\ Lett.\  B {\bf 724} (2013) 92
 [arXiv:1305.4846 [hep-ph]];
  S.~Antusch, C.~Gross, V.~Maurer and C.~Sluka,
  Nucl.\ Phys.\ B {\bf 877} (2013) 772
  [arXiv:1305.6612 [hep-ph]];
  S.~Antusch, C.~Gross, V.~Maurer and C.~Sluka,
  Nucl.\ Phys.\ B {\bf 879} (2014) 19
  [arXiv:1306.3984 [hep-ph];
  S.~F.~King,
  JHEP {\bf 1401} (2014) 119
  [arXiv:1311.3295 [hep-ph]];
    S.~F.~King,
  JHEP {\bf 1408} (2014) 130
  [arXiv:1406.7005 [hep-ph]].


\bibitem{SO10}
    I.~de Medeiros Varzielas, S.~F.~King and G.~G.~Ross,
  Phys.\ Lett.\ B {\bf 648} (2007) 201
  [hep-ph/0607045];
%
    S.~F.~King and M.~Malinsky,  
  Phys.\ Lett.\ B {\bf 645} (2007) 351
  [hep-ph/0610250];
  %
    F.~Bazzocchi and I.~de Medeiros Varzielas,
  Phys.\ Rev.\ D {\bf 79} (2009) 093001
  [arXiv:0902.3250 [hep-ph]];
%
B.~Dutta, Y.~Mimura and R.~N.~Mohapatra,
  JHEP {\bf 1005} (2010) 034
  doi:10.1007/JHEP05(2010)034
  [arXiv:0911.2242 [hep-ph]];
%
    K.~M.~Patel,
  Phys.\ Lett.\ B {\bf 695} (2011) 225
  [arXiv:1008.5061];
%
  P.~S.~Bhupal Dev, R.~N.~Mohapatra and M.~Severson,
  Phys.\ Rev.\ D {\bf 84} (2011) 053005
  [arXiv:1107.2378];
%
  P.~S.~Bhupal Dev, B.~Dutta, R.~N.~Mohapatra and M.~Severson,
  Phys.\ Rev.\ D {\bf 86} (2012) 035002
  [arXiv:1202.4012];
%
  I.~de Medeiros Varzielas and G.~G.~Ross,
  JHEP {\bf 1212} (2012) 041
  [arXiv:1203.6636 [hep-ph]];
   A.~Anandakrishnan, S.~Raby and A.~Wingerter,
  Phys.\ Rev.\ D {\bf 87} (2013) 5,  055005
  [arXiv:1212.0542 [hep-ph]].

  \bibitem{Froggatt:1978nt}
  C.~D.~Froggatt and H.~B.~Nielsen,
  Nucl.\ Phys.\ B {\bf 147} (1979) 277.
 
\bibitem{deMedeirosVarzielas:2008en}
  I.~de Medeiros Varzielas,
  arXiv:0804.0015 [hep-ph].

\bibitem{Varzielas:2015aua}
  I.~de Medeiros Varzielas,
  JHEP {\bf 1508} (2015) 157
  [arXiv:1507.00338 [hep-ph]].
   
\bibitem{Ibanez:1982fr}
  L.~E.~Ibanez and G.~G.~Ross,
  Phys.\ Lett.\ B {\bf 110} (1982) 215;
  B.~R.~Greene, K.~H.~Kirklin, P.~J.~Miron and G.~G.~Ross,
  Nucl.\ Phys.\ B {\bf 292} (1987) 606;
  I.~de Medeiros Varzielas and G.~G.~Ross,
  hep-ph/0612220;
  L.~E.~Ibanez and G.~G.~Ross,
  Comptes Rendus Physique {\bf 8} (2007) 1013
  [hep-ph/0702046 [HEP-PH]].
  I.~de~Medeiros~Varzielas,
  arXiv:0801.2775 [hep-ph];
  R.~Howl and S.~F.~King,
  Phys.\ Lett.\ B {\bf 687} (2010) 355
  [arXiv:0908.2067 [hep-ph]].

\bibitem{Murayama:1994tc}
  H.~Murayama and D.~B.~Kaplan,
  Phys.\ Lett.\ B {\bf 336} (1994) 221
  doi:10.1016/0370-2693(94)90242-9
  [hep-ph/9406423].
         
\bibitem{Beringer:1900zz}
  J.~Beringer {\it et al.}  [Particle Data Group Collaboration],
  Phys.\ Rev.\ D {\bf 86} (2012) 010001.

\bibitem{Antusch:2013jca}
  S.~Antusch and V.~ Maurer,
 JHEP \textbf{1311} (2013) 115
  [arXiv:1306.6879 [hep-ph]].

\bibitem{Gonzalez-Garcia:2014bfa}
  M.~C.~Gonzalez-Garcia, M.~Maltoni and T.~Schwetz,
  JHEP {\bf 1411} (2014) 052
  [arXiv:1409.5439 [hep-ph]].

\bibitem{Ade:2015xua}
  P.~A.~R.~Ade {\it et al.}  [Planck Collaboration],
  arXiv:1502.01589 [astro-ph.CO];
  R.~H.~Cyburt, B.~D.~Fields, K.~A.~Olive and T.~H.~Yeh,
  arXiv:1505.01076 [astro-ph.CO].


 

\end{thebibliography}
\end{document}